\documentclass[acmsmall]{acmart}

\AtBeginDocument{%
  }

\newcommand{\Foutse}[1]{\textcolor{teal}{{\it [Foutse: #1]}}}

\newcommand{\fltool}{\textit{FL4Deep}}

\usepackage{tikz}
\newcommand*\circled[1]{\tikz[baseline=(char.base)]{
            \node[shape=circle,draw,inner sep=2pt] (char) {#1};}}
            
\usepackage{listings}
\usepackage{subfig}
\usepackage[most]{tcolorbox}
\usepackage{color, colortbl}
\usepackage{multirow}
\usepackage{adjustbox}
\usepackage{makecell}
\usepackage{booktabs}
\usepackage{inputenc}
\usepackage{nicematrix}
\usepackage{lipsum}
\usepackage{color}
\usepackage{algorithm}
\usepackage{algpseudocodex}

\definecolor{codegreen}{rgb}{0,0.6,0}
\definecolor{codegray}{rgb}{0.5,0.5,0.5}
\definecolor{codepurple}{rgb}{0.58,0,0.82}
\definecolor{backcolour}{rgb}{0.95,0.95,0.92}

\lstdefinestyle{mystyle}{
    backgroundcolor=\color{backcolour},
    commentstyle=\color{codegreen},
    keywordstyle=\color{blue},
    numberstyle=\tiny\color{codegray},
    stringstyle=\color{codepurple},
    basicstyle=\ttfamily\footnotesize,
    breakatwhitespace=false,         
    breaklines=true,                 
    captionpos=b,                    
    keepspaces=true,                 
    numbers=left,                    
    numbersep=5pt,                  
    showspaces=false,                
    showstringspaces=false,
    showtabs=false,                  
    tabsize=2
}

\newcommand{\Hilight}{\makebox[0pt][l]{\color{cyan!35}\rule[-4pt]{\linewidth}{12pt}}}

\setcopyright{acmlicensed}
\copyrightyear{2024}
\acmYear{2024}
\acmDOI{XXXXXXX.XXXXXXX}

\acmJournal{TOSEM}
\begin{document}

    % This work was supported by: Fonds de Recherche du Québec (FRQ), the Canadian Institute for Advanced Research (CIFAR) as well as the DEEL project CRDPJ 537462-18 funded by the National Science and Engineering Research Council of Canada (NSERC) and the Consortium for Research and Innovation in Aerospace in Québec (CRIAQ), together with its industrial partners Thales Canada inc, Bell Textron Canada Limited, CAE inc and Bombardier inc.

\title[Fault Localization in DL-based Software]{Fault Localization in Deep Learning-based Software: A System-level Approach
% \Amin{Approach?}
}

\author{Mohammad Mehdi Morovati}
\email{mehdi.morovati@polymtl.ca}
\orcid{1234-5678-9012}
\affiliation{%
  \institution{SWAT Lab., Polytechnique Montréal}
  \city{Montréal}
  \state{Quebec}
  \country{Canada}
}

\author{Amin Nikanjam}
\email{amin.nikanjam@polymtl.ca}
\affiliation{%
  \institution{SWAT Lab., Polytechnique Montréal}
  \city{Montréal}
  \state{Quebec}
  \country{Canada}
}

\author{Foutse Khomh}
\email{foutse.khomh@polymtl.ca}
\affiliation{%
  \institution{SWAT Lab., Polytechnique Montréal}
  \city{Montréal}
  \state{Quebec}
  \country{Canada}
}

% \affiliation{%

% \affiliation{%

% \affiliation{%

% \affiliation{%

% \affiliation{%

% \affiliation{%

% \affiliation{%

\renewcommand{\shortauthors}{Morovati et al.}

\newcommand{\Amin}[1]{\textcolor{blue}{{\it [Amin: #1]}}}

\begin{abstract}
  Over the past decade, 
  Deep Learning (DL)
  has become an integral part of our daily lives, with its widespread adoption in various fields, particularly in safety-critical domains. This surge in 
  DL usage has heightened the need for developing reliable DL 
  software systems, making Software Reliability Engineering (SRE) techniques essential. 
  Given that fault localization is a critical task in SRE, researchers have proposed several fault localization techniques for DL-based software systems, primarily focusing on faults within the DL model. While the DL model is central to DL components, there are numerous other elements that significantly impact the performance of DL components. As a result, fault localization methods that concentrate solely on the DL model overlook a large portion of the system. To address this, we introduce \fltool{}, a system-level fault localization technique based on a Knowledge Graph (KG). For the first time, \fltool{} considers the entire DL development pipeline to effectively localize faults across the DL-based systems.
  \fltool{} first extracts the necessary static and dynamic information from DL software systems, then generates a KG by analyzing the collected information. 
  Finally, it provides a ranked list of potential faults by inferring relationships from the KG. In an evaluation using 100 faulty DL scripts, \fltool{} outperformed four previous approaches in terms of accuracy for three out of six DL-related faults, including issues related to data (84\%), mismatched libraries between training and deployment (100\%), and loss function (69\%). Additionally, \fltool{} demonstrated superior precision and recall in fault localization for five categories of faults including three mentioned fault types in terms of accuracy, plus issues related to the insufficient training iteration with $0.89$ and $0.62$ and activation function with $0.89$ and $0.92$ for precision and recall, respectively. 
  Sensitivity analysis of \fltool{} components also indicates that static information has the most significant impact on the performance of \fltool{}.
  
\end{abstract}





\maketitle

\section{Introduction}
Nowadays, 
Deep Learning (DL) has
become an integral part of our everyday life. DL has been used in extensive applications across diverse domains including Natural Language Processing (NLP)~\cite{khurana2023natural}, autonomous driving~\cite{yurtsever2020survey}, financial~\cite{henrique2019literature}, and medical systems~\cite{bhattacharya2021deep}. These software systems which employ DL components are called DL-based systems~\cite{morovati2024bug}. With the increasing dependence of current software systems on DL components, it is crucial to ensure the reliability of these components. 
DL-based systems, similar to traditional software systems, are prone to a variety of software faults~\cite{humbatova2020taxonomy}. 
One of the most critical tasks in ensuring the reliability of software systems (either DL-based or traditional) is debugging, which focuses on detecting and fixing faults~\cite{ieee8016712}. Even when faults are known to exist due to the system's faulty behavior, the process of finding the location of faults' root causes remains a significant challenge~\cite{wong2016survey}. 
Fault localization, which plays a key role in debugging, involves identifying the specific location of faults' root causes within the system and fixing them~\cite{alipour2012automated}.

In traditional software systems, %expected behavior is defined statically. 
when a mismatch occurs between the expected and actual outputs, it indicates the presence of a fault in the software system~\cite{ammann2016introduction}. Similarly, faults in DL software refer to discrepancies between the program's current behavior and the expected outcome~\cite{zhang2020machine}.
% \Amin{true, but is it always true? \Mehdi{I think so, Do you have another idea?}}. 
However, unlike traditional software where the logic is represented through control flow coded by developers,
the output in DL-based software is determined by a 
trained model~\cite{wardat2021deeplocalize}. As a result, 
DL-based systems introduce new challenges that cause more complex debugging than traditional software~\cite{islam2020repairing}. 
Consequently, fault detection and localization in DL-based systems are more difficult than in traditional systems. 
For instance, when a classifier produces an incorrect classification, it does not necessarily indicate a fault in the DL-based systems. 
%In fact, due to the randomness of DL, it cannot guarantee 100\% classification accuracy~\cite{zhang2020machine}.
Although researchers have increasingly focused on developing testing and debugging techniques for DL-based systems, fault localization has received comparatively less attention~\cite{wardat2022deepdiagnosis}.
This is expected
due to the distinct challenges inherent in fault localization within the context of DL-based systems.
Compared to traditional software, 
the root causes of faults
in DL-based software are more varied, located in three main components: 1) the DL program code, 2) the DL framework, and 3) the data used to train the DL models~\cite{tambon2021silent,islam2019comprehensive}. In this study, we focus specifically on faults within the DL software system code, 
% scripts
% \Amin{script?I don't understand}, 
excluding issues related to DL frameworks and data.

\begin{lstlisting}[language=Python, float, floatplacement=H, label=fig:motive_sample,caption=SO post (\href{https://stackoverflow.com/questions/42081257}{\#42081257}) showing faults in loss function (line 15) and metrics (line 16) that pose a challenge for developers to identify and address.
% \Foutse{please highlight what people should be focusing on in this listing...for example the lines where there is a fault!!! And make the connection to them in the text!}
]
model.add(embedding_layer)
model.add(Dropout(0.25))
# convolution layers
model.add(Conv1D(nb_filter=32, filter_length=4, 
            border_mode='valid', activation='relu'))
model.add(MaxPooling1D(pool_length=2))
# dense layers
model.add(Flatten())
model.add(Dense(256))
model.add(Dropout(0.25))
model.add(Activation('relu'))
# output layer
model.add(Dense(len(class_id_index)))
model.add(Activation('softmax'))
model.compile(loss='binary_crossentropy', 
                optimizer='adam', metrics=['accuracy'])
\end{lstlisting}

% \Amin{this is very clear! I would summarize the parts we have prior to it!}
The fundamental differences between DL-based and traditional software paradigms introduce novel types of faults unique to DL-based systems. As a result, debugging techniques developed for traditional software cannot be efficiently applied to DL-based systems~\cite{humbatova2021deepcrime,morovati2024bug}.
% \Amin{add citations to back the claim}. 
% These differences result in generating new types of faults in ML-based software systems~\cite{morovati2023bugs}. 
For example, Listing \ref{fig:motive_sample} shows a script from a SO post discussing a DL-based system encountering `bad performance' (\href{https://stackoverflow.com/questions/42081257}{\#42081257}). The developer who submitted the post indicated that she has been unable to detect and localize the issue, specifically faults within the \emph{loss function} and \emph{metrics} parameters (Lines 15 and 16 of Listing \ref{fig:motive_sample})
% \Foutse{please indicate the faulty lines in your listing!}.
% using the existing fault localization techniques. 
These faults are mostly identified through manual code review which is time-consuming and needs DL expertise~\cite{nikanjam2021automatic}. 
Since fault localization is one of the most time-consuming aspects of debugging, it significantly impacts the overall effort required for software debugging~\cite{wong2016survey}. 
That is, inaccurate fault localization can mislead the debugging process, resulting in wasted time and effort for developers. 
Therefore, providing fault localization techniques for DL-based systems that consider essential features of these systems can significantly assist DL-based systems' developers. As illustrated in Fig.~\ref{fig:ml_based_system}, the DL component consists of various elements. While the DL model is the central element, it represents only a small part of the whole DL components~\cite{sculley2015hidden}. Therefore, faults within the DL component can originate from any of these elements, not just the DL model itself~\cite{chen2020comprehensive}. Therefore, efficient fault localization for DL-based systems should target not only the DL model itself but also the entire training pipeline and its associated components.  %However, existing fault localization techniques for DL-based systems focus solely on the DL model and its training phase, overlooking other critical elements.

% That is, we need to provide new debugging methodologies for ML-based systems that consider essential features of these systems. 

% \Amin{in this paragraph, you need to provide more details about your approach}

To address this gap, we propose a system-level fault localization technique namely \fltool{}, which builds upon Information Retrieval (IR)-based 
and history-based fault localization approaches. \fltool{} targets 
% addresses\Foutse{targets?} 
DL faults across all components of the DL pipeline, including data, the DL model and its training, and components dealing with the DL-based system deployment process.
% \Foutse{what do you mean by "system deployment"? it is vague! what specifically to the approach targets in the deployment setting?}. 
To identify and localize faults, \fltool{} extracts both static and dynamic information from DL-based systems and constructs a Knowledge Graph (KG) representing the system's features such as dataset features used to train DL models, model hyperparameters, used environment to train models, etc.
% \Foutse{can you briefly describe the information contained in that knowledge graph?}. 
Additionally, it generates an ordered list of potential faults and their root causes, inferred through a set of rules designed based on the commonly known faults in DL-based systems.
% \Foutse{predefined how? what rationale guided the rules design?}. 
Evaluation results show that \fltool{} outperforms other fault localization approaches (including \textit{DeepFD}~\cite{cao2022deepfd}, \textit{AutoTrainer}~\cite{zhang2021autotrainer}, \textit{DeepLocalize}~\cite{wardat2021deeplocalize}, and \textit{UMLAUT}~\cite{schoop2021umlaut}) 
% \Foutse{which one? can you list them?} 
in 83\% of the 
% \Foutse{what do you mean by ''review faults cases'?} 
% reviewed fault cases, 
100 buggy samples used for their comparison,
in terms of precision and recall.
To summarize, this research makes the following contributions:
\begin{itemize}
    \item We present the first technique that analyzes the entire pipeline of deep learning-based system development to effectively localize faults. %the bugs.
    \item We highlight the key challenges in fault localization of DL-based systems. %\Foutse{the term demonstrate is used for proof and may be too strong here!}
    \item We present fault localization techniques originally developed for traditional software systems, which can be adapted to best suit the unique characteristics of deep learning models.
    % \Foutse{not clear what this means!}
    % most fitted fault localization methods with DL nature. 
    \item We provide a dataset of real-world buggy DL codes extracted from SO posts and GitHub repositories.
    \item We release the source code of \fltool{} alongside our datasets to facilitate its use by other researchers~\cite{replication_package_fl4ml}.
\end{itemize}

\textbf{The remaining of this paper is structured as follows.}
The main related studies are reviewed in Section \ref{sec:related_works}.
Section \ref{sec:background} provides background information on fault localization in DL-based systems. In Section \ref{sec:method}, we present \fltool{} and its methodology in detail. %a detailed introduction to . 
Section~\ref{sec:result} presents the results and analysis of the
comparison of \fltool{} with four existing fault localization methods for DL-based systems. 
% In Section~\ref{sec:result}, we discuss the result and analysis regarding \fltool{}.
% we explore fault localization techniques originally developed for traditional software systems and discuss their potential adaptation for DL-based systems.
%Section \ref{sec:result} presents the results of comparing \fltool{} with four previous fault localization methods for DL-based systems. In Section \ref{sec:discuss}, 
%we discuss fault localization techniques originally developed for traditional software systems that could be adapted for DL-based systems. 
Section \ref{sec:threats} outlines 
% \Foutse{what do 'potential' means here!}potential
threats to the validity of \fltool{}. Finally, we conclude the paper and outline future research directions in Section \ref{sec:conclusion}.

\section{Related works}
\label{sec:related_works}
% \Amin{maybe it is better if we move related works (since it reviews the approaches you examined) to the beginning of the paper, like after background. See it with Foutse}
In this section, we review techniques developed for localizing faults in DL-based systems. 
% \subsection{DEBAR}
\textit{DEBAR} is a static analysis tool that detects numerical bugs at the architecture level of DL programs~\cite{zhang2020detecting}. \textit{DEBAR} uses two main categories of abstracting methods to detect numerical faults~\cite{cousot1977abstract} including 1) tensor and 2) numerical values. Regarding tensor abstracting methods, \textit{DEBAR} uses array expansion, array smashing, and tensor partitioning abstracting methods. Furthermore, DEBAR uses interval and affine relation analysis abstraction techniques as numerical abstraction methods. 
Numerical faults refer to the issues represented as \textit{`NaN'}, \textit{`INF'}, or crashes during the training phase. 
It also checks the program source codes for the most common unsafe operations such as \textit{Exp}, \textit{Log}, etc.
Results of evaluating \textit{DEBAR} show that it outperforms other existing static numerical fault detection techniques, in terms of accuracy (93.0\%) without decreasing the performance. 

% \subsection{DeepLocalize}
\textit{DeepLocalize} introduces a white box-based technique to localize faults in DL programs 
% (based on Keras~\cite{chollet2018keras}) 
using two basic steps~\cite{wardat2021deeplocalize}. 
The first step involves generating intermediate code from the source code of the DL program. This intermediate code is created to verify whether the DL statements are identifiable.
% observable \Foutse{what does 'observable mean here exactly?}. 
In the subsequent step, \textit{DeepLocalize} employs a white-box method to dynamically analyze the traces generated during model training.
% which analyses traces of model training dynamically.  
% The second approach, provides a customized callback function collecting the information of ML program during its training phase and analyzing it to identify possible bugs in ML model. 
\textit{DeepLocalze} detects the faulty layers or hyperparameters leading to bugs in DL program~\cite{wardat2021deeplocalize}. 
It is worth noting that \textit{DeepLocalize} requires both the source code of the DL software system and the model training logs.
That is, it needs an executable DL program without any compile error to be able to analyze it and find possible bugs. 
To evaluate \textit{DeepLocalize}, 40 buggy samples in total
% \Foutse{is it 40 in total or 40 from each source? please be precise!} 
were collected from GitHub and SO, showing that \textit{DeepLocalize} detects bugs and their root cause in 34 and 21 out of 40, respectively.
% It is also evaluated by 
% Results show that DeepLocalize detects bugs and their root cause in 34 and 21 out of 40, respectively.  

% The tool also could not report the line of code related to the bug. Indeed, it reports the number of layers that deal with identified bugs. This tool also needs the runnable DL program without any compile error to be able to analyze it and find out possible bugs. 
% \subsection{Umlaut}
\textit{UMLAUT}, the Usable Machine LeArning Utility, is a tool that helps DL developers identify, understand, and debug DL programs 
% provided based on Keras
~\cite{schoop2021umlaut}. 
\textit{UMLAUT} integrates with the DL program to collect model training information and heuristically assess the model's structure and behavior. After analyzing the gathered data, Umlaut identifies potential issues, representing them as error messages and offering best practice solutions in the form of code snippets. \textit{UMLAUT} provides a Keras callback function that enables the capture of model training details. Based on the severity of the identified issue, Umlaut classifies the messages into three levels: warning, error, and critical.
To recommend best practices for addressing these issues, \textit{UMLAUT} draws from various sources, including lecture notes~\cite{conciseML}, books~\cite{goodfellow2016deep}, and expert blogs~\cite{Recipe4NN}. \textit{UMLAUT} claims to detect issues across multiple stages of the DL pipeline, such as data preparation, where it identifies problems like input data exceeding typical limits, NaN values in loss or input data, shape mismatches in image inputs, and unexpected validation accuracy. It also detects model architecture issues, such as missing activation functions, the absence of a softmax layer, and the use of multiple activation functions in the final layer. Furthermore, Umlaut flags parameter tuning issues, including suboptimal learning rates, potential overfitting, and excessively high dropout rates. By providing these insights, Umlaut helps optimize the DL model development process.
% \subsection{NauraLint}

\textit{NauraLint}, a model-based fault detection approach for ML
programs provide a technique to detect faults in the DL programs using meta modeling and graph transformation~\cite{nikanjam2021automatic}. 
To analyze a DL program, \textit{NeuraLint} translates it into a model, based on the provided meta-model for DL programs. Next, it uses a model-based verification to check 23 rules. The rules have been inferred from 1) research papers studying bugs in DL programs, 
2) public datasets of faulty DL programs, and 
3) official tutorials of DL frameworks. 
\textit{Neuralint} was evaluated using 34 faulty ML/DL programs extracted from GitHub and SO.
The results show a recall of 70.5\% and a precision of 100\% when detecting %precision of the provided methodology in detecting 
bugs and design issues. 
% Results represent 70.5\% recall and 100\% precision as evaluation of the provided methodology in detecting bugs and design issues. 
% \Foutse{please also add a description of DeepChecker!!! https://dl.acm.org/doi/10.1145/3529318}
% \subsection{DRLinter}

Braiek et al.\cite{ben2023testing} introduced a property-based debugging approach called \textit{TheDeepChecker}, aimed at identifying 17 deep learning (DL) bugs previously documented by Humbatova et al.\cite{humbatova2020taxonomy}. To achieve this, they extracted key features for each DL component (such as initial random parameters and output activation functions) and developed principles based on these features to detect DL bugs.
To debug a DL program, \textit{TheDeepChecker} begins by collecting dynamic information during model training, such as hidden activations, predictions, and losses. It then applies various statistical calculations to reduce the dimensionality of the extracted data. In the following step, \textit{TheDeepChecker} flags layers where more than half of the neurons have died, as determined by a threshold in the outputs. Next, it evaluates the DL program against critical values and erroneous behaviors using pessimistic boundaries. 
Finally, an approximative component assesses the results from the previous steps against an anticipated set of possible states and behaviors to identify bugs within DL programs.
To evaluate \textit{TheDeepChecker}, the authors tested it on real-world DL buggy samples sourced from SO and GitHub. The comparison between \textit{TheDeepChecker} and \textit{Amazon SMD} showed that \textit{TheDeepChecker} outperformed \textit{Amazon SMD}, achieving 75\% accuracy in bug identification compared to \textit{Amazon SMD}'s 60\%.

\textit{DRLinter}, a model-based fault detection technique for deep reinforcement learning (DRL), is a tool to discover bugs in DRL programs~\cite{nikanjam2022faults}. 
Reinforcement Learning (RL) is a subcategory of ML, where the main goal is to achieve maximum reward. 
% In RL, an agent receives the information of the environment and make the optimal decision based on it, over time~\cite{lapan2018deep}. 
DRL refers to the integration of the DL methodology into RL approaches to improve sequential decision-making~\cite{goodfellow2016deep}. 
In the first step, \textit{DRLinter} transforms the DRL program into a graph. Next, it uses a generic meta-model for DRL programs to check 11 defined rules and validate them against possible faults. 
% DRLinter provided 11 different rules to check them and make sure that there is no fault in the DRL program. 
\textit{DRLinter} has been evaluated using 15 synthetic DRL programs, where errors were artificially injected into the source code, and 6 real-world DRL programs. The results show that \textit{DRLinter} successfully detects faults in all synthetic examples. In the real-world samples, it achieved a recall of 75\% and a precision of 100\% in identifying faults. 

% \subsection{Autotrainer}
\textit{AutoTrainer} is an automated system for detecting and repairing training issues in deep neural networks. It identifies five distinct types of training faults in Convolutional Neural Networks (CNNs) and Recurrent Neural Networks (RNNs)~\cite{zhang2021autotrainer}, helping to optimize the training process. 
CNN is known as a kind of Neural Network (NN) using convolution in at least one of the network layers. 
CNN is mostly used for analyzing a grid of values (e.g., images)~\cite{goodfellow2016deep}. 
RNN is also considered a subcategory of NN used for analyzing sequential data. 
It is important to take into account that connections in RNN can create a cycle where the output of some nodes may be the subsequent input of the same nodes~\cite{graves2013speech}. 
\textit{AutoTrainer} collects and analyzes model training logs to detect possible training faults. 
As \textit{AutoTrainer} focuses on the model training phase and its related faults, it can detect 1) vanishing gradient, 2) exploding
% \Foutse{exploding?} 
gradient, 3) dying ReLU, 4) oscillating loss, and 5) slow convergence problems. 
\textit{AutoTrainer} also provides a built-in solution to fix any of the identified faults, ensuring prompt correction when issues are detected.
%the fault that can be used in case of finding any of the mentioned faults.
% In case Autotrainer detects any of the mentioned faults in the model, 
% it uses its built-in solutions to fix the fault. 
Evaluation results show that \textit{AutoTrainer} identified 316 faults in 262 reviewed DL programs and successfully fixed 309 of them. Additionally, the average accuracy of the fixed programs improved by up to 1.5 times.
%The results of the Autotrainer evaluation show that it discovers 316 faults in 262 reviewed DL programs and fixes 309 of the detected faults. Besides, the average accuracy improvement of the fixed programs has been improved up to 1.5 times. 

% \subsection{DeepFD}

\textit{DeepFD} is a learning-based fault diagnosis and localization technique for DL programs ~\cite{cao2022deepfd}. \textit{DeepFD} collects and analyzes data from model training to identify potential faults within DL programs. It extracts 20 distinct features, such as loss and accuracy at each epoch, to detect faults in DL programs. It employs multi-label versions of three classifiers—K-Nearest Neighbors (KNN), decision tree, and random forest—to diagnose these faults.
Subsequently, the source code of the DL program is converted into an Abstract Syntax Tree (AST) to localize the faults identified in the previous step.
Finally, \textit{DeepFD} reports several suspicious lines of code that are likely to be the root causes of the detected faults.

%DeepFD, a learning-based fault diagnosis and localization technique, is a tool to address the limitations of fault localization in DL programs as a learning problem~\cite{cao2022deepfd}. 
%It gathers and analyzes information from model training to locate the possible faults in DL programs. 
%It extracts 20 different features (such as loss and accuracy in every epoch) from the model training to detect faults in DL programs. DeepFD uses multi-label versions of three classifiers including K-Nearest Neighbors (KNN), decision tree, and random forest to diagnose faults in DL programs.Next, the DL program source code is transferred into the Abstract Syntax Tree (AST) to localize the faults that are detected in the previous step. Finally, it reports several suspicious lines of code that can be the root causes of discovered faults. 

% \subsection{DeepDiagnosis}
\textit{DeepDiagnosis} is a tool that recognizes faults, reports symptoms of faults, localizes the found faults, and provides suggestions for fixing them~\cite{wardat2022deepdiagnosis}. 
Similar to the previous approaches, \textit{DeepDiagnosis} also collects information from model training and analyzes it to detect possible faults. 
To this end, it provides a callback function that should be passed to the model training method. 
Next, it analyzes the gathered data to detect 8 different %types of faults; %eone of the 8 different 
fault symptoms;
including saturated activation, exploding tensor, accuracy not increasing, dead node, loss not decreasing, unchanged weight, exploding gradient, and vanishing gradient. 
% \Foutse{can you list the others?}. %and their related root causes. 
In the next step, it uses a decision tree to localize the root cause of the identified faults and formulate recommendations for their correction. Several changes to the model structure can be recommended by the tool to fix the identified faults; 
% i.e.,  %leading to 7 recommended changes in the model structure such as 
change loss/activation function, change optimizer, change weight/bias initialization, change learning rate, change number of training layers, change batch size, and change size of training data.
% \Foutse{can you list the others?} 
\textit{DeepDiagnosis} was evaluated on a total of 444 ML programs, sourced from GitHub, Stack Overflow, or generated by \textit{AutoTrainer}~\cite{zhang2021autotrainer}. 
The comparison of \textit{DeepDiagnosis} with \textit{UMLAUT}~\cite{schoop2021umlaut}, \textit{DeepLocalize}~\cite{wardat2021deeplocalize}, and \textit{AutoTrainer}~\cite{zhang2021autotrainer} approaches, based on 56 buggy models from GitHub and SO, demonstrated that \textit{DeepDiagnosis} outperformed all the other approaches. However, its accuracy was lower than all mentioned approaches when applied to the samples generated by \textit{AutoTrainer}.
% \Foutse{all the other methods mentioned above? can you be precise?}, 

% \subsection{Deep4Deep}
Wardat et al.~\cite{wardat2023effective} introduced an approach called \textit{Deep4Deep} to automatically debug DNN programs and localize faults by mapping extracted model features to specific model problems. They implemented a Long Short-Term Memory (LSTM) model to learn the relationship between symptoms of a faulty model and their root causes. The LSTM model captures patterns of model issues using features extracted from the DNN model.
\textit{Deep4Deep} leverages both dynamic and static information from DNN models. Dynamic information includes model parameters (e.g., weights, metrics) observed during training, such as loss, activation functions, data range, and the vanishing gradient. Static information is derived directly from the DNN model's source code, independent of execution.
Additionally, Wardat et al. compared \textit{Deep4Deep} with existing methods, including UMLAUT~\cite{schoop2021umlaut}, DeepLocalize~\cite{wardat2021deeplocalize}, Autotrainer~\cite{zhang2021autotrainer}, and DeepDiagnose~\cite{wardat2022deepdiagnosis}. Their results showed that \textit{Deep4Deep} outperforms these methods in terms of fault detection and localization accuracy, time efficiency, and the clarity of information provided to users regarding faults and their root causes.

%Wardat et al.~\cite{wardat2023effective} introduced an approach, namely \textit{Deep4Deep}, to automatically debug DNN programs and localize their faults, by mapping extracted model features onto the problems of models. 
%To this end, they implemented a Long Short-Term Memory (LSTM) model to learn the relationship between the symptoms of the buggy model and their root causes. In other words, LSTM model learns the patterns of model problems, using features extracted from the model. Generally, \textit{Deep4Deep} uses both dynamic and static information of the DNN models. 
%DNN models' dynamic information refers to the values of model parameters (weights and metrics) during model training (e.g. model loss and activation functions, data range, vanishing gradient, etc). Static information of DNN models also means the information achieved from the source code of DNN models, without running them. Moreover, they compared their proposed approach with other existing methods (UMLAUT~\cite{schoop2021umlaut}, DeepLocalize~\cite{wardat2021deeplocalize}, Autotrainer~\cite{zhang2021autotrainer}, and DeepDiagnose~\cite{wardat2022deepdiagnosis}) and reported that \textit{Deep4Deep} outperforms others, in terms of fault detection and localization accuracy, consumed time, and presented information regarding the faults and their root causes to the users.

\section{Background}
\label{sec:background}
\subsection{Software Debugging: Error, Fault, and Failure}
In the software community, a fault is considered the manifestation of a software error resulting in an incorrect software functionality~\cite{ieee5733835:2010}. Software error is a programmer mistake that can be grammatical (a problem in one or more lines of code) or logical (a problem in satisfying one or more software requirements)~\cite{galin2004software}. Generally, all software errors may not become a software fault. That is, an erroneous line of code is converted to the fault and affects the software functionality when it is executed. If a user tries to use a faulty section of the software and activates a software fault, it can lead to software failure. Software failure also refers to the inability of software to perform required functionality~\cite{riccio2020testing}.

%\subsection{}
Software Quality Assurance (SQA) is a systematic approach involving essential tasks designed to ensure confidence that a software system will meet its technical requirements~\cite{galin2004software}. 
% \Mehdi{What is software quality attributes?}
% To manage SQA in more efficient way, software requirement has been classified into different groups namely software quality attributes~\cite{NDUKWE2023111524}. 
It is widely accepted that among all software quality attributes, software reliability is the most significant one, where each attribute assesses the conformance level of the system with identified requirements~\cite{lyu2007software,morovati2023bugs}. Software Reliability Engineering (SRE) is the methodology to make sure that the operation of software during a specific period is failure-free~\cite{RADJENOVIC20131397}. Fault removal is one of the most important SRE approaches aiming at finding and removing existing faults. 
Fault removal techniques use validation and verification approaches which are known as software testing techniques~\cite{galin2004software}.

Software testing is the first SQA tool to verify the expected behavior of a software unit, several integrated software units, or the whole software system, before its installation on the customer side environment~\cite{galin2004software,ieee5733835:2010}. 
The main role of software testing is to find and address software defects that have negative effects on the software quality, using a set of test cases. 
Each test case includes a set of test data, execution conditions, and expected results to ensure that the developed software complies with requirements~\cite{ieee5733835:2010}. When test cases reveal that software behavior deviates from expected outcomes, the development team initiates a software debugging process to identify and correct the underlying errors~\cite{myers2011art}.

% \subsection{Machine Learning (ML) and Deep Learning (DL)}
\subsection{DL-based Software Systems}
Machine Learning (ML), a subfield of Artificial Intelligence (AI), involves algorithms that learn from data to create intelligent computer programs~\cite{mccarthy2007artificial}. Generally, ML algorithms improve their performance over time by leveraging past experiences~\cite{bell2020machine}. Deep Learning (DL), a branch of ML, utilizes neural networks with a large number of layers~\cite{amershi2019software}. DL is particularly well-suited for tasks involving large and complex datasets~\cite{goodfellow2016deep}. By increasing the number of layers and units per layer, DL models can represent more complex functions and solve more complicated problems.

\begin{figure}[!t]
    \centering
    \includegraphics[width=\columnwidth]{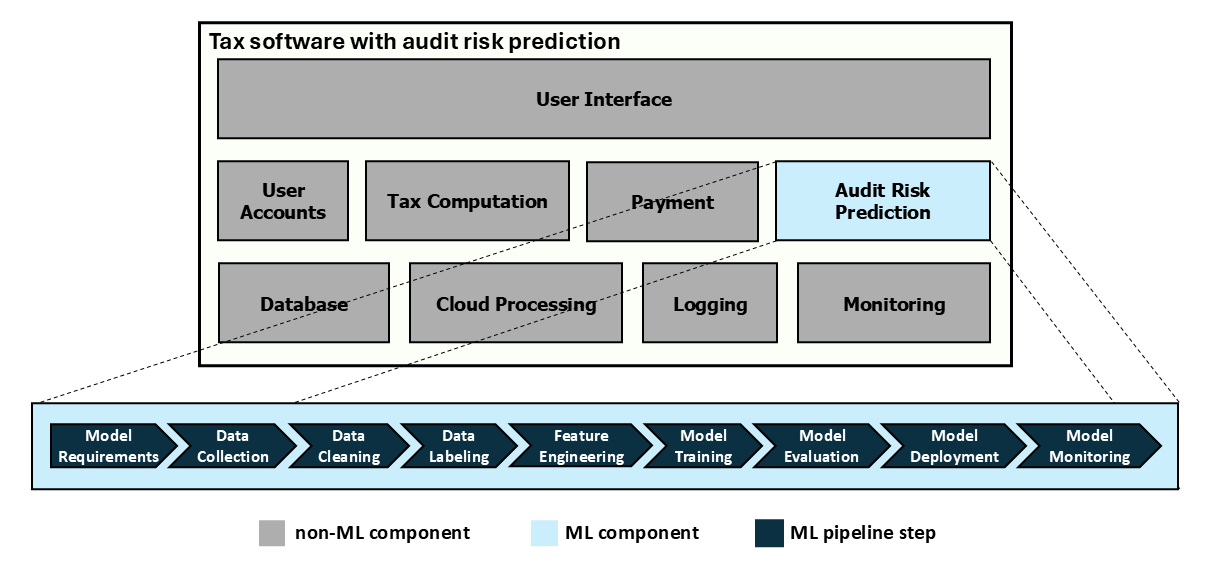}
    \caption{High-level view of a DL-based tax software system 
    % (inspired from \cite{sculley2015hidden})
    }
    \Description{DL-based online transcription system}
    \label{fig:ml_based_system}
\end{figure}

% \subsection{ML-based Systems}
A software system is a system consisting of one or more software components~\cite{ieee5733835:2010}. A software component is a well-defined entity with an independent structure representing a set of functions~\cite{iso2010information}. A DL-based system is a software system including at least one DL component~\cite{morovati2023bugs}. Accordingly, a DL component is a software component whose functionality relies on a deep neural network. Fig.~\ref{fig:ml_based_system} shows a high-level view of a DL-based online transaction system. 
% \begin{figure}[!t]
%     \centering
%     \includegraphics[width=\columnwidth]{figures/ml_pipeline.pdf}
%     \caption{Pipeline of ML-based system development process (adopted from~\cite{amershi2019software})}
%     \Description{Pipeline of ML-based system}
%     \label{fig:ml_pipeline}
% \end{figure}
It also represents the general pipeline of the DL-based development process, comprising nine distinct steps. In the model requirement phase, system designers assess the feasibility of implementing various system features using DL and select the most suitable models. Data collection involves integrating available datasets or creating a dataset specific to the software system. In the data cleaning step, inaccurate and noisy data are filtered out to ensure a high-quality dataset. Data labeling refers to assigning informative labels to each record in the dataset. Feature engineering encompasses the activities involved in selecting appropriate features for the DL models. Model training involves training the selected DL models using the chosen features and the cleaned data. During the model evaluation phase, the trained model is tested using a separate dataset and evaluated based on predefined metrics (e.g., accuracy). Model deployment refers to deploying the trained and evaluated models into the target environments and devices. Finally, in the model monitoring phase, models deployed in real-world environments are continuously monitored to detect potential system bugs or performance issues.

\subsection{Error, Fault, and Failure in DL-based Systems}
The concept of faults and their localization differs significantly between traditional and ML/DL programs due to their fundamental characteristics~\cite{wardat2021deeplocalize}. In traditional software systems, the expected behavior is typically defined statically, and a mismatch between expected and actual outputs indicates the presence of a bug~\cite{ammann2016introduction}. Bugs in ML/DL programs also arise from discrepancies between the current behavior and the expected behavior~\cite{zhang2020machine}. However, ML/DL bugs can originate from three sources: the program code, the DL framework, and the data used~\cite{tambon2021silent, islam2019comprehensive}. This study focuses specifically on bugs within the ML/DL program code, excluding issues related to the DL frameworks and data.
ML/DL programs are more difficult to debug in comparison %is more complex compared 
to traditional software~\cite{islam2020repairing}. For instance, when a classifier produces an incorrect classification, it is not necessarily indicative of a bug in the DL program, as ML models cannot guarantee 100\% accuracy due to their statistical nature~\cite{zhang2020machine}. Moreover, unlike traditional software, where logic is represented as control flow, DL programs rely on the weights between neurons to determine the output~\cite{wardat2021deeplocalize}.

According to the IEEE Standard Glossary of Software Engineering Terminology, a software fault is defined as a static defect in the software~\cite{ieee5733835:2010}. Software faults are generally classified into two main categories: 1) functional faults, which occur when the software fails to meet functional requirements, and 2) non-functional faults, which involve issues in the methodologies used to fulfill those requirements~\cite{morovati2023bugs}. Similarly, a fault in DL programs is considered a deficiency in their behavior~\cite{21448:IEEESafety}.

\subsection{Fault Localization}
Generally speaking, fault localization is considered as an operation that receives a faulty program and a set of test cases as input and generates a ranked list of suspicious program elements~\cite{zou2019empirical}.
Program elements can be taken into account in three levels including 1) statement (line of code), 2) method, and 3) file. 
% We can categorize 
Fault localization techniques have been categorized into two main groups, based on the methodology used for the analysis~\cite{wong2023handbook,ernst2003static}. 
\begin{itemize}
    \item \textbf{\textit{Static:}} Methods that rely solely on the source code, without executing the application, extract necessary information directly from the available code and analyze it to identify suspicious elements.
    % that don't need to run the application and just use source code of the program. 
    \item \textbf{\textit{Dynamic:}}Techniques that require running the program and utilizing runtime information (e.g., stack traces, bug reports, etc.). These techniques try to monitor the system execution during the testing process and collect necessary dynamic data. Dynamic analysis techniques may also extract static information from the available source code to enrich the collected dynamic data for analysis.  
\end{itemize}

% From another perspective, we can categorize fault localization techniques based on the failure type that maybe identified. For instance, crash which is known as the easiest type of system failure to be detected by fault localization techniques, 
% can be discovered just by analysing stack trace.

% \subsubsection{Fault Localization Techniques}
% \label{subsec:fault_localizing_tech}

% In this section, we examine the primary fault localization techniques that have been developed for traditional software systems. 
Various techniques have been developed for fault localization in software systems, each bringing a unique approach with specific strengths and limitations. The following sections present an overview of these methods.
\subsubsection{Spectrum-based Fault Localization (SBFL)}
A program spectrum refers to a measurement of a program's runtime behavior (such as code coverage). Spectrum-based fault localization leverages these runtime behavior measurements of test cases to identify and locate faults~\cite{ARRIETA201818}. In other words, this approach analyzes the dynamic behavior of the program during the execution of a test suite to determine how likely each statement contains a fault~\cite{10.1145/2000791.2000795}.
% \Mehdi{If we consider the training logs (callback function outputs) as a program spectrum, we can categorize all existing ML/DL fault localization techniques in this group}
\subsubsection{Mutation-based Fault Localization (MBFL)}
Mutation analysis is the assessment process of the test suite's effectiveness in detecting various types of faults. Generally, each mutant replaces an operand or expression with another, thereby altering a statement~\cite{7985698}. 
Mutation testing is a software testing technique that involves introducing faults into the program under test (referred to as mutants) and analyzing the differences in behavior between the mutants and the original program~\cite{petrovic2021does}. 
Mutation-based fault localization identifies suspicious mutants and uses this information to pinpoint the location of faulty statements~\cite{10.1002/stvr.1509}. 
Suspiciousness level of a program statement is determined by how frequently it affects both passed and failed tests.

\subsubsection{Fault Localization Using Program Slicing}
As software systems have grown larger and more complex, software debugging has become increasingly challenging. Program slicing is a debugging technique
that narrows the focus by
% \Foutse{what exactly?} 
isolating program statements relevant to a specific computation, effectively removing irrelevant parts and transforming the program into a minimal form~\cite{harman2001overview}. 
A program slice represents a subset of the program that influences a particular system behavior. In other words, it is the portion of the program that affects the values of specific statements of interest~\cite{xu2005brief}. 
Program slicing is generally divided into two main types: static and dynamic.
Static slicing relies solely on information available statically, without executing the program. In contrast, dynamic slicing considers only the statements executed for a particular input, resulting in a smaller, more focused slice compared to static slicing~\cite{zhang2007study}.
% semantic aspect
% \Foutse{what do you mean by 'semantic aspect'?}. \Foutse{slicing consists of finding the parts of a program relevant to the value of a chosen set of variables at some chosen point in a program. A slice is constructed by deleting the parts of the program that are irrelevant to those values.}
% \Foutse{executable statements for any input?}potential executed statements for any input
In bug localization, program slicing simplifies the program by extracting a minimal subset of the program that directly influences the incorrect behavior, making it easier to investigate. The next step is to identify the specific statements within this slice that are responsible for the faulty behavior~\cite{soremekun2021locating}.
% \Foutse{this is vague! slicing is a comment technique, please use precise information about it!} and highlighting potential problematic lines of code for further investigation.

\subsubsection{Fault Localization Using Stack Trace Analysis}
A stack trace is a sequence of active stack frames generated during program execution, offering crucial information for debugging~\cite{zou2019empirical}. 
Each function call generates a stack frame that remains active until the function returns. 
Fault localization methodologies based on stack trace gather active traces of passed and failed system execution. 
Next, the fault localizer tries to identify the active functions at the point of a system crash and localize the fault, by analyzing these stack traces~\cite{gong2014locating}.
It is worth mentioning that this technique is mostly used to check the reason for program crashes.
% \Mehdi{For ML/DL applications, we can use customized callback function to collect all detailed logs during training phase}
\subsubsection{Predicate Switching Fault Localization}
A predicate controls the execution of a program by determining its branching paths. Predicate switching involves running a program through various control flows~\cite{zhang2006locating}. 
If altering the outcome of a condition expression causes a test case to pass instead of fail, the predicate is considered critical and may be the root cause of the failure. 
The predicate-switching fault localization technique uses mutations to modify predicate conditions and evaluates the results of the program's execution.
\subsubsection{Information Retrieval-based Fault Localization}
Information Retrieval (IR) is a methodology used to extract relevant information from large collections of unstructured data~\cite{manning2010introduction}, primarily employed for text indexing and searching in documents. 
In the context of fault localization, IR-based techniques take bug reports as input to generate a ranked list of files likely related to the reported issue. 
Notably, this approach does not require program execution information (e.g., test case results) and relies solely on the content of the bug reports~\cite{wong2016survey}.
% \Mehdi{It is usable just when we have a bug report. In fact, in case the ML application has a problem and does not make any error message, we can not use this technique.}
\subsubsection{History-based Fault Localization}
History-based fault localization uses development history information to localize bugs, operating on the premise that source files with a history of a higher number of bugs are more likely to contain defects in the future. 
Put simply, history-based methods require a substantial amount of processed historical data~\cite{Venkat_history_fl}. 
In the history-based fault localization approach, program elements are ranked according to their likelihood of being defective, similar to the process used in bug prediction~\cite{rahman2011bugcache}. 
% \Mehdi{It seems this approach is usable just in a single project. That is, we can not use the history information of one software system for another one. }

\subsubsection{Learning to Rank Fault Localization}
With the significant increase in computational power over the past decade, DL has emerged as one of the most popular fields in computer science. DL techniques have also been used to enhance fault localization techniques. %, similar to its impact on other domains.
The Learning to rank Fault Localization techniques train DL models to prioritize files by their likelihood of containing defects, leveraging suspiciousness scores derived from various SBFL formulas~\cite{yu2019empirical}. 

\subsection{Most Fitted Fault Localization Techniques for DL-based systems}
In traditional software systems, there are several approaches to localize faults. We discuss here their applicability to DL-based software systems.
Spectrum-Based Fault Localization (SBFL)~\cite{10.1145/2000791.2000795} relies on analyzing the dynamic behavior of the software during execution. However, as highlighted in the previous section, the inherent stochasticity of DL systems poses challenges for generating test cases, a known issue referred to as the oracle problem in DL testing~\cite{seca2021review}. 
This challenge significantly impacts the effectiveness of SBFL.
Learning-to-rank fault localization techniques, which build on SBFL, also encounter similar issues due to the reliance on the dynamic behavior of the system and test case generation.
Mutation-Based Fault Localization (MBFL)~\cite{10.1002/stvr.1509} faces challenges related to the time required for fault localization. Given that model training (the core aspect of DL-based systems) is a time-intensive process, this can severely hinder the efficiency of MBFL approaches, where we need to run applications against various mutants. 
Techniques such as predicate switching and fault localization through program slicing, which rely on program code or control flow analysis, are not well-suited for DL-based systems~\cite{usman2023overview}. This is because all lines of code within DL components are typically executed during each run of the DL software system, making these techniques less effective. 
Moreover, fault localization methods based on stack trace analysis may be inadequate for DL-based systems, as they often lack access to low-level traces from DL framework operations.

In contrast, history-based fault localization techniques present fewer challenges. These approaches analyze the probability of specific faults being linked to various elements of the application by examining the history of fault occurrences, making them more adaptable to DL-based systems.
Information Retrieval (IR)-based techniques may also encounter fewer difficulties compared to others when applied to DL-based systems, in case they work based on the static information of the DL-based systems. The challenges of extracting relevant static information from DL software systems can be managed without significantly affecting the efficiency of the fault localization process.

\fltool{} provides a fault localization approach for DL-based systems by combining IR-based and history-based techniques, both of which are well-suited to the unique characteristics of DL-based systems. This approach enhances the accuracy and performance of the fault localization process for DL-based systems.

\subsection{Challenges in Fault Localization of DL-based Systems}
Due to the specific challenges in testing and debugging DL-based systems, fault localization in these systems also presents unique difficulties. One major challenge is the reliance on test cases, which are fundamental to traditional software fault localization~\cite{zou2019empirical}. 
However, the DL testing community widely accepts that generating effective test cases for DL software systems is particularly challenging~\cite{sugali2021software}.
This difficulty stems from the inherent stochasticity in DL systems, which introduces uncertainty in their outputs. Additionally, the design of training data is crucial for both the performance and accuracy of DL models. Consequently, creating accurate test cases for DL software systems requires expertise not only in ML but also in related fields like data management~\cite{gao2019ai}.
DL frameworks (such as \textit{Keras}, \textit{TensorFlow}, and \textit{PyTorch}), which are designed to simplify the development of DL-based systems, play a significant role in modern ML development~\cite{zhang2020machine}. However, creating test cases based on these frameworks can be particularly challenging due to internal faults~\cite{rivera2021challenge,tambon2021silent}, 
rapid changes in their APIs~\cite{yang2022comprehensive,haryono2021characterization}, and alterations in their internal operations. For example, Islam et al.~\cite{islam2020repairing}
% \Amin{isn't this reference too old?} 
reported that approximately 26\% of \textit{TensorFlow} operations were modified between versions 1.10 and 2.0.
Beyond the previously mentioned challenges, the inherent randomness and uncertainty in DL can lead to different results with each execution of a DL software system~\cite{zhang2020machine}. To address this issue, some fault localization techniques for DL-based systems involve running the application multiple times. For instance, DeepFD~\cite{cao2022deepfd} localizes faults by analyzing data from 10 runs of the application. However, since model training is a highly time-consuming process, this approach can significantly impact the performance of the fault localization technique.  

\fltool{} tackles these challenges by leveraging both static and dynamic information, which reduces the impact of randomness in the fault localization process and eliminates the need to execute DL-based systems multiple times to localize faults. Furthermore, \fltool{} is implemented using \textit{Keras 2.8} and \textit{TensorFlow 2.8}, ensuring compatibility with all \textit{Keras} and \textit{TensorFlow} versions in the $2.x$ series.

\begin{figure}[!t]
    \centering
\includegraphics[width=0.6\linewidth]{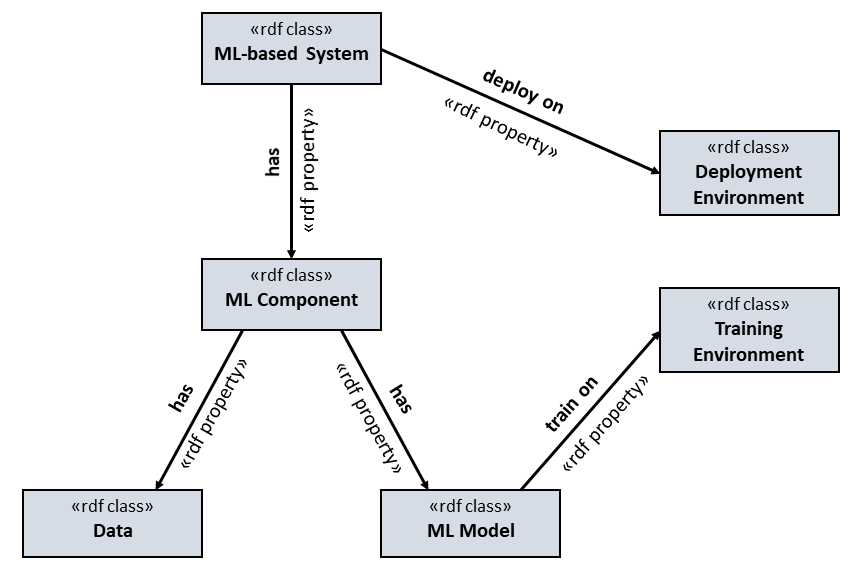}
    \caption{Sample KG of a DL-based system}
    \label{fig:kg}
\end{figure}

\subsection{Knowledge Graph (KG)}

A Knowledge Graph (KG) is both a specialized type of Knowledge Base (KB) system and a form of labeled, directed graph that organizes and represents large amounts of structured knowledge. KGs are designed to store entities and their relationships in a way that allows for efficient querying, reasoning, and extraction of meaningful insights~\cite{chen2020review}.
%KG includes sets of interlinked facts that describe events or entities and relationships among them. 
Google started the development of KG in 2012 to develop a structure 
including important aspects of human knowledge that can be found in data sources~\cite{googleKG2015}. 
KG can organize knowledge from various information sources effectively to represent knowledge about certain domains~\cite{fensel2020knowledge}. 
Since KG provides a contextualized understanding of
data, it has received more attention in recent years~\cite{barrasa2023building}. 
A key feature of KG in data representation is that they treat the links between facts (the edges in the KG) as equally important as the facts themselves. Figure \ref{fig:kg} represents a simple KG of a DL-based system. 

% \subsection{Resource Description Framework (RDF)}

The Resource Description Framework (RDF) is a standard model for representing information about resources~\cite{rdf_w3}, which is the core building block of KG. 
RDF is designed to enable efficient information processing by applications, going beyond merely presenting data in a human-readable format. By representing relationships between resources, RDF is particularly well-suited for building interlinked datasets~~\cite{pan2009resource}.
%RDF is designed to facilitate the processing of information by applications, rather than only presenting it in a form that is understandable to humans. Because RDF enables users to express relationships among resources, it is often considered an excellent choice for implementing interlinked data~\cite{pan2009resource}. 
In RDF, linked data is represented as triples consisting of a \textit{subject}, \textit{predicate}, and \textit{object}. The \textit{subject} and \textit{object} denote entities, while the \textit{predicate} specifies the relationship between entities.

% \subsection{Notation 3 (N3)}

Notation3 (N3) is a logical programming language considered a superset of RDF~\cite{notation3}. 
N3 extends RDF's capabilities by enhancing its representational power and enabling decision-making through data manipulation, information access, and reasoning. 
It allows developers to make statements about entities (including logical implications) which is known as declarative programming. 
Consequently, an N3 reasoning engine can infer new information from the declared statements, making it a suitable option for automating decision-making processes or enriching KG~\cite{n3Language}. 

% \subsection{Link Prediction in KG}

\begin{figure}[!t]
  \centering
  \frame{
  \subfloat[]{\includegraphics[width=0.305\textwidth]{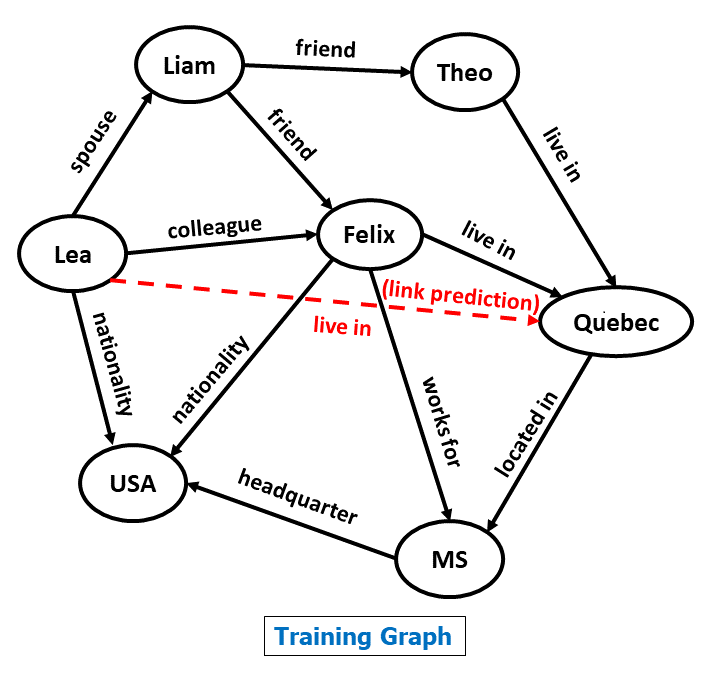}\label{fig:lp_transductive}}}
  \frame{
  \subfloat[]{\includegraphics[width=0.62\textwidth]{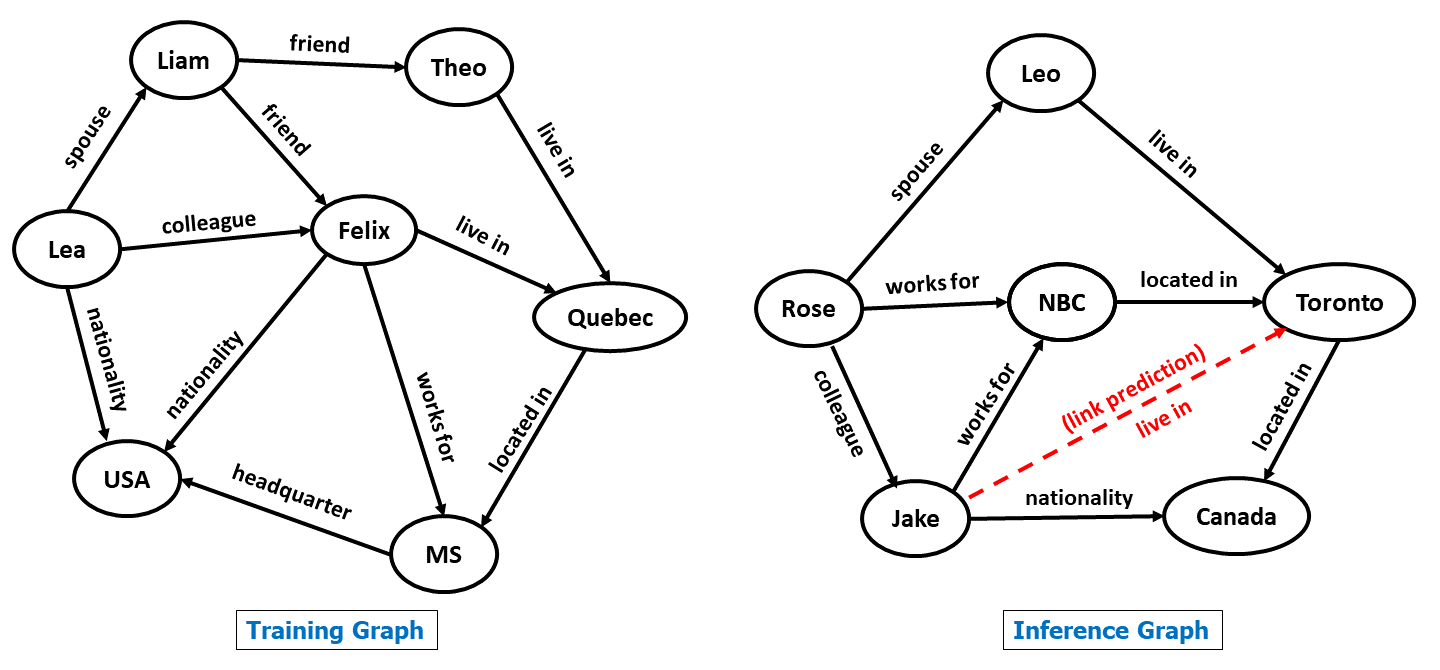}\label{fig:kp_inductive}}
  }
  \caption{Link prediction in KG based on the type of inference (a) transductive link prediction, (b) inductive link prediction.
  }
  \label{fig:KG_LP}
  \vspace{-1em}
\end{figure}

Despite recent advances in KG, even the largest KGs remain incomplete. For instance, FreeBase~\cite{bollacker2008freebase}, one of the largest public domain KGs, has over 70\% of its person entities lacking information about their place of birth~\cite{shirvani2023comprehensive,west2014knowledge}. 
To address this issue, link prediction has been developed to infer missing relationships among entities~\cite{rossi2021knowledge}.
Traditionally, link prediction 
% \Amin{what is "it"? our KG or any KG?}
has relied on various heuristic metrics based on the paths between graph nodes, such as Katz\cite{katz1953new} and PageRank~\cite{page1999pagerank}. However, recent advancements in DL inspired researchers to develop new techniques for link prediction in KGs using Graph Neural Networks (GNNs)~\cite{schlichtkrull2018modeling,vashishth2019composition}. At the same time, several datasets have also been created specifically for link prediction tasks in graphs, including FB15K~\cite{bollacker2008freebase} and WN18~\cite{bordes2013translating}. 
There are generally two main techniques for link prediction in KGs, based on the type of inference: transductive and inductive~\cite{galkin2022open}. Transductive link prediction involves training the DL model on the same graph used for inference~\cite{baek2020learning}, meaning that it uses known entities to predict links between them. 
In contrast, inductive link prediction involves using a different graph for training than the one used for inference~\cite{galkin2021nodepiece}, enabling the prediction of possible links between previously unseen entities. Fig.\ref{fig:KG_LP} shows the difference between transductive and inductive link prediction.

\section{\fltool{}}
\label{sec:method}

\begin{figure}[!b]
    \centering
    \includegraphics[width=\linewidth]{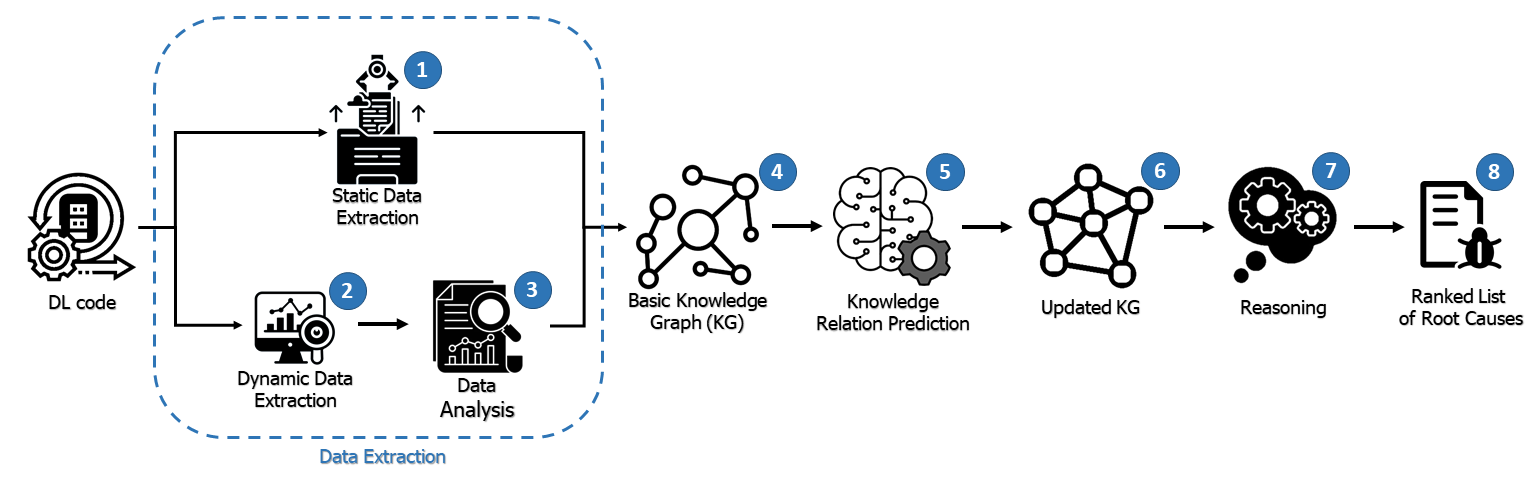}
    \caption{High-level view of \fltool{} methodology}
    \label{fig:method_fl4ml}
\end{figure}

This section outlines the pipeline of our proposed fault localization technique, \fltool{}. Fig. \ref{fig:method_fl4ml} provides a high-level view of the methodology we followed in \fltool{}. Algorithm \ref{alg:fl4ml} represents the pseudocode of the \fltool{} algorithm.

\begin{algorithm}[t]
\caption{\fltool{} Algorithm}
\label{alg:fl4ml}
\renewcommand{\algorithmicrequire}{\textbf{Input:}}
\renewcommand{\algorithmicensure}{\textbf{Output:}}
\begin{algorithmic}[]
\small
\rmfamily
\Require Dataset, DL code, training environment, deployment environment
\Ensure Ranked list of fault's root causes

\State $SI \gets$ staticInfoExt$(Dataset, DL code, trainingEnvironment)$
\newline \color{black!75} \ttfamily \hspace{1cm} // Extracting static information

\rmfamily
\color{black}  
\State $DI \gets$ dynamicInfoExt$(modelTrainingLogs)$
\newline \color{black!75} \ttfamily \hspace{1cm} // Extracting dynamic information 

\rmfamily
\color{black}
\State $KG \gets$ KgGenerator$(SI, DI)$  
\newline \color{black!75} \ttfamily \hspace{1cm} // Creating a KG based on the collected information 

\rmfamily
\color{black} 
\State $RCs \gets$ reasoningEngine$(KG)$ 
\newline \color{black!75} \ttfamily \hspace{1cm} // Localizing faults using a reasoning engine 

\rmfamily
\color{black} 
\State $RankedRCs \gets$ ranking$(RCs)$
\newline \color{black!75} \ttfamily \hspace{1cm} // Ranking the identified faults 

\end{algorithmic}
\end{algorithm}

\subsection{Dataset Preparation}
\label{subsec:dataset}
In this study, we prepare two datasets of buggy DL code, referred to as the training and validation datasets. The training dataset consists of 75 real-world buggy DL programs, extracted from SO and GitHub repositories. We used this dataset to train the ML models employed in two components of \fltool{} 
represented as \circled{3} and \circled{5} in Fig. \ref{fig:method_fl4ml}. 
% \Amin{be clear, mention those ml models and components}. 
Of 75 samples.
% these\Foutse{these what?}
17 buggy codes were sourced from the research conducted by Cao et al.\cite{cao2022deepfd}, and 30 were extracted from the \textit{Defect4ML} dataset\cite{morovati2023bugs}. Besides, 12 additional buggy samples were reported by both sources. For the 16 remaining buggy samples, we utilized the Stack Exchange Data Explorer\footnote{https://data.stackexchange.com/stackoverflow/query}, a portal providing an up-to-date database of SO posts. We extracted posts tagged with both \textit{machine-learning'}/\textit{deep-learning'} and \textit{keras'}/\textit{tensorflow'}.
The first two authors randomly sampled a set of posts for manual review, focusing on those with accepted answers; that included DL code snippets in the questions; and were reproducible. Each selected post was then labeled based on the faults identified in the accepted answers on SO or discussed in the corresponding GitHub Pull Requests (PRs). 
%The two first authors then randomly reviewed these posts, selecting those with accepted answers, containing DL code snippets in the questions, and being reproducible. Each sample is labeled according to the faults identified in the accepted answers from SO posts or discussions within the GitHub Pull Requests (PR).

The validation dataset includes 100 buggy DL scripts, used to compare the performance of \fltool{} with previous fault localization techniques for DL-based software systems. 
To create this dataset, we utilized 20 SO posts gathered from prior studies~\cite{cao2022deepfd,wardat2021deeplocalize}. Additionally, we applied various mutation operators specifically designed for DL software systems (such as modifying the loss function and changing the activation function~\cite{chen2020comprehensive,humbatova2021deepcrime}) to these 20 samples to generate new buggy samples. 
% \Amin{from what? I mean you applied mutations on which samples?}. 
%For labeling the validation dataset samples, we used the reported faults from the answers in SO posts for those originating from SO. For samples generated through mutation operators, we assigned labels based on the faults injected by the respective mutation operators. All samples included in both the training and validation datasets are available in the replication package of this study
For labeling the validation dataset samples, we used the reported faults from SO posts for those derived from SO. For samples generated using mutation operators, we assigned labels based on the faults introduced by the corresponding mutation operators. All samples in both the training and validation datasets are available in the replication package accompanying this study~\cite{replication_package_fl4ml}.

\subsection{Extracting Required Information from DL codes}

\begin{figure}[!b]
    \centering
    \includegraphics[width=\columnwidth]{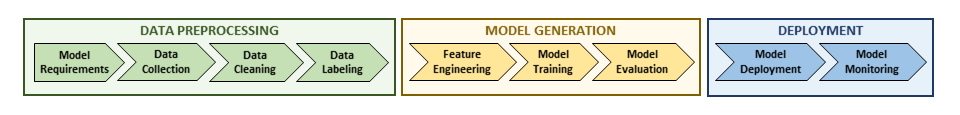}
    \caption{Pipeline of DL-based system development process divided into three parts (adopted from~\cite{amershi2019software})}
    \Description{Pipeline of DL-based system development}
    \label{fig:ml_pipeline_divided}
\end{figure}

To extract information from DL software systems from the software under test,
% \Amin{from the software under test?}, 
we divide the entire DL-based system pipeline into three modules: 
1) data preprocessing, 
2) model generation, and 
3) system deployment. Fig.~\ref{fig:ml_pipeline_divided} illustrates how the various stages of the DL-based system development pipeline are categorized into these three modules. Algorithm \ref{alg:fl4ml_data_extract} presents the pseudocode of \fltool{} information extraction.

\begin{algorithm}[t]
\caption{\fltool{}'s data extraction 
% \Foutse{add comments to explain the statements}
}\label{alg:fl4ml_data_extract}
\renewcommand{\algorithmicrequire}{\textbf{Input:}}
\renewcommand{\algorithmicensure}{\textbf{Output:}}
\begin{algorithmic}[]

\small
\rmfamily

\Require Dataset, DL code, training environment, deployment environment
\Ensure static and dynamic info regarding DL code

\State $S1 \gets Analysis(dataset)$ 
\newline \color{black!75} \ttfamily \hspace{1cm} // Extracting info regarding dataset

\color{black} \rmfamily
\State $S2 \gets Analysis(trainingEnvironment)$
\newline \color{black!75} \ttfamily \hspace{1cm} // Extracting info regarding training environment

\color{black} \rmfamily
\State $S3 \gets modelHyperparameters$
\newline \color{black!75} \ttfamily \hspace{1cm}  // Extracting info regarding model hyperparameters

\color{black} \rmfamily
\Repeat
    \If{end-of-training-epoch}
        \State $DI.append(modelTrainingLogs)$
        \newline \color{black!75} \ttfamily \hspace{2cm} // Extracting dynamic info after each training epoch
        
        \color{black} \rmfamily 
    \EndIf
\Until{end-model-training}
\State $S4 \gets Analysis(deploymentEnvironment)$
\newline \color{black!75} \ttfamily \hspace{1cm}  // Extracting info regarding deployment environment

\color{black} \rmfamily 
\State $SI \gets combine(S1, S2, S3, S4) $

\end{algorithmic}
\end{algorithm}

\fltool{} utilizes both static and dynamic information
% \Amin{isn't it static and dynamic analysis?}\Amin{in the rest you have definitions of such information, I suggest to add them here as well} 
(presented as \circled{1} and \circled{2} in Fig.\ref{fig:method_fl4ml}, respectively) of DL-based systems to identify and localize faults. Static information refers to information within the DL software system
% that can be extracted from the source code of an DL application, 
which is extracted from the source code before running the DL 
% which remains consistent across all executions of the DL
software system (such as the size of the dataset, DL model structure and its layers, activation functions, optimizer, etc)~\cite{kumar2010survey}. 
% \Amin{did you get this definition from somewhere? I meant "static" since it is a technical term in software\Mehdi{I added a reference for it, but it does not mention it the same as what I said here}}.
% \Amin{will you discuss those components? if not, just say that we collect different information}. 
% To parse DL codes and gather necessary static information, \fltool{} 
% takes various parts of the DL system as input (such as the dataset, DL model, training environment, and deployment environment) and collects the required information. As an example, dataset is used as the input of a component to extract information regarding data preprocessing module. 
To gather necessary static information from DL-based systems, \fltool{} takes various DL system components (including the dataset, DL model, training environment, and deployment environment) as input and collects the required information. 
% \Foutse{why only this example on data preprocessing data? one wonders how other static data from the DL-based system are collected!} 
% To extract static information regarding data, the dataset serves as input to a component collecting details such as train/test ratio, number of samples in the train/test part, etc. 
% Information regarding training and deployment environment (such as version of \textit{Python}, version of installed libraries, etc) is gathered before model training and before executing the DL-based systems in the deployed environment, respectively. Moreover, static information regarding the DL model such as model hyperparameters, layers, etc is extracted exactly after creating the model, but before starting model training.

To collect the necessary static information from DL-based systems, \fltool{} takes various components of the system(including the dataset, DL model, training environment, and deployment environment) as input. For dataset-related information, \fltool{} extracts details like the train/test split ratio, the number of samples in each part, and more. Information about the training and deployment environments, such as the \textit{Python} version and installed libraries, is gathered before model training and before executing the system in the deployed environment, respectively. 
Additionally, static information about the DL model, such as hyperparameters and layer configurations, is collected immediately after the model is created but before training begins.

% that extracts details about the data preprocessing module.
% \Amin{it's unclear to me! I might be wrong, but can you put a list of those static information that you extract?}
 
% Additionally, we provide a Keras callback function to capture detailed information about the DL model (including its structure, layers, activation functions, optimizer, etc.).
% \Amin{I think you mean: to add the callback to the code, so extract the information? if so, clarify, and add an example}

Dynamic information is collected during the training of DL-based software which can vary depending on runtime conditions~\cite{kumar2010survey}. 
% \Amin{any references to cite?}.
In \fltool{}, we gather model training logs as dynamic information (presented as \circled{2} in Fig.\ref{fig:method_fl4ml}). 
% In our proposed methodology\Amin{\fltool{}?}, we gather system logs\Amin{which logs? it's unclear} during model training as dynamic information. 
% To achieve this, we use Keras callback implemented to extract a part of static information. 
% in the previous step
% \Amin{by previous you mean for static data extraction? if so why we have it there, explain it here} 
% In other words, we use Keras callback functionality to 
\fltool{} extracts neuron weights, accuracy, loss, validation accuracy, and validation loss after each epoch of model training as the dynamic information.
Similar to previous studies on fault localization in DL-based systems using dynamic information~\cite{cao2022deepfd}, \fltool{} processes the extracted dynamic information using eight statistical operators, as shown in Table~\ref{tab:statistical_operators}.
% \Amin{this is clear! for static part, it isn't\Mehdi{No, it is done just for dynamic information}}. 
These operators help capture diagnostic features from the gathered data. For instance, the standard error of the mean indicates how much the sample mean would vary if the study were repeated with new samples from the same population~\cite{lee2015standard}. 

\begin{figure}[!t]
    \centering
    \includegraphics[width=0.75\linewidth]{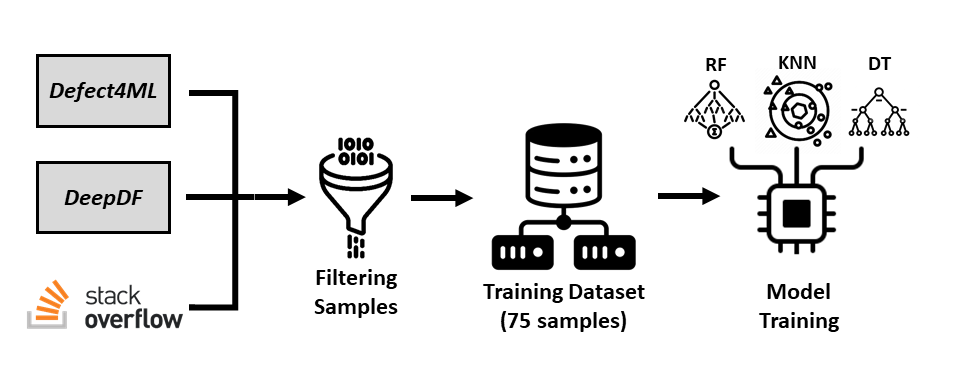}
    \caption{Process of creating training dataset and training Random Forest (RF), Decision Tree (DT), and K-nearest Neighbors (KNN) models}
    \label{fig:dynamic_analysis}
\end{figure}

In the next step, we employ three widely used ML models, i.e., Random Forest (RF)\cite{rigatti2017random}, Decision Tree\cite{charbuty2021classification}, and K-Nearest Neighbors (KNN)~\cite{zhang2021challenges}, to analyze dynamic information and predict potential faults (presented as \circled{3} in Fig.\ref{fig:method_fl4ml}). 
These models are trained using the training dataset presented in subsection~\ref{subsec:dataset}.
% \Foutse{what dataset exactly?}.
% \Amin{"the training dataset" you mean?\Mehdi{Yes, exactly. Should I mention it in another way?}}.
% of 75 buggy DL codes. 
% The results of applying statistical operations to the dynamic information extracted from the 75 training samples are used as input for model training, while the corresponding labels (fault types) serve as the output. 
% That is, we train each model as a classifier that aims to predict potential faults based on the extracted information. 
The results of applying statistical operations to the dynamic information extracted from the 75 training samples are used as input for model training, with the corresponding fault types serving as the output labels. 
That is, each model is designed as a classifier aiming to predict potential faults based on the extracted data.
Fig. \ref{fig:dynamic_analysis} represents the process to train these three models. 
% \Foutse{can you add a figure showing the dataset used for this training? it is not very clear from your description!}.
% \Amin{as the goal is predicting a potential fault per extracted information?}.
The DL faults that can be identified through dynamic analysis include issues related to the loss function, activation function, optimizer, insufficient iteration, and inappropriate learning rate
% \Foutse{how do you identify exactly? do you formulate a classification task for which these types of faults are to be predicted? give details please}. 
These faults are considered the most common issue regarding the model training phase of DL-based software systems~\cite{cao2022deepfd}.
% \Amin{citation?}.
Since a DL software system 
% \Amin{be consistent! sometimes you have DL software system, and other time DL application. we need to have one term, unless you clarify that you are using them interchangeably}
may encounter multiple faults simultaneously, we use the multi-class versions of these models~\cite{vens2008decision,wu2016ml,zhang2005k}. After obtaining the results from all three models, we apply a majority voting process to aggregate 
% determine\Amin{aggregate?} 
the final prediction of potential faults. Algorithm \ref{alg:fl4ml_dynamic} shows the pseudocode of dynamic information analysis.
Listing \ref{list_dynamic_sample} shows a code snippet from the SO post \href{https://stackoverflow.com/questions/51181393}{\#51181393}. After extracting dynamic information and applying statistical operations, the result is used as input for all three models. 
In this case, both the RF and KNN models flagged an issue with the learning rate, while the decision tree model reported no issue. Therefore, through majority voting, the final output identifies a learning rate issue.

\begin{lstlisting}[language=Python, float, floatplacement=H, label=list_dynamic_sample,
caption=An example of predicting the fault (too small learning rate) by analyzing dynamic information extracted during model training.
]
data = data = linspace(1,2,100).reshape(-1,1)
y = data*5
model = Sequential()
model.add(Dense(1, activation = 'linear', input_dim = 1))
model.compile(optimizer = 'rmsprop', loss = 'mean_squared_error', metrics = ['accuracy'])
model.fit(data,y,epochs =200,batch_size = 32)
\end{lstlisting}

% \Foutse{some concrete examples of models, inputs, outputs, and aggregation would he helpful!}

% The features from buggy DL applications serve as inputs, while the corresponding bugs are used as labels for training these models. The labels used for training these models include faults related to loss function, activation function, optimizer, insufficient iteration, and inappropriate learning rate, as these are the most common fault types in DL-based systems~\cite{cao2022deepfd}.

\begin{algorithm}[!b]
\caption{Prediction of faults based on the dynamic information}\label{alg:fl4ml_dynamic}
\renewcommand{\algorithmicrequire}{\textbf{Input:}}
\renewcommand{\algorithmicensure}{\textbf{Output:}}
\begin{algorithmic}[]

\small
\rmfamily

\Require Extracted dynamic information
\Ensure Dynamic Information analysis

\State $data \gets$ Process$(dynamicInfo)$ 
\newline \color{black!75} \ttfamily \hspace{1cm} // Processing dynamic info using statistical operators

\color{black}\rmfamily
\State $Models \gets [RF, DT, KNN]$ 
\newline \color{black!75} \ttfamily \hspace{1cm} // models including Random Forest, Decision Tree, and KNN

\color{black}\rmfamily
\State $faults \gets \{\}$
\For{$model$ in $Models$}

        \State $F \gets predict(model, data)$ 
        \newline \color{black!75} \ttfamily \hspace{1.5cm} // predicting faults using processed dynamic info
        
        \color{black}\rmfamily
        \State $faults.insert(F)$

\EndFor

\State $predictedFaults \gets$ majorityVoting$(faults)$
\newline \color{black!75} \ttfamily \hspace{1cm} // concluding the faults using majority voting 

\end{algorithmic}
\vspace{1em}
\end{algorithm}

% To train these models, we use 69 
% % trained models with 58 
% buggy samples introduced by Cao et al.~\cite{cao2022deepfd}, and \textit{Defect4ML}~\cite{morovati2023bugs}.
% These samples are derived from SO posts and issues raised in GitHub repositories. 
% Following previous research, we use features from 10 executions of each sample to train the models~\cite{cao2022deepfd}.

% \Amin{both analysis leads to facts, but it is unclear from your description. say that we have 2 sources to gather infor about bugs: static and dynamic. for static we do this and for dynamic we do that ... and we collect some facts this way. and mention examples to make it clear for the reader}

\subsection{Creating 
% and Reasoning 
% \Foutse{what do you mean by 'creating and reasoning'?}
KG}
% The main reason behind using KG in \fltool{} 
% where relationships among nodes play a central role in our approach.
In this step, \fltool{} constructs a KG to detect faults and localize their root causes. Algorithm \ref{alg:fl4ml_faults_localize} represents the pseudocode for constructing the KG and utilizing it to localize faults within the system.
%of KG creation and fault localization based on that. 
The list of the faults identifiable by \fltool{}, along with a brief description of each fault is provided in Table~\ref{tab:fl4ml_rules_reasoning}. 
These faults have been collected from previous studies on various components of DL software systems, such as data~\cite{joseph2022optimal,tan2021critical}, model training~\cite{cao2022deepfd,nikanjam2021automatic,nwankpa2018activation,glorot2010understanding}, and deployment~\cite{chen2020comprehensive,morovati2023bugs}, as well as Q\&A forums (e.g., SO posts). 
These faults are presented in the KG as rules. 
To generate the KG, \fltool{} firstly incorporates  
the collected static information and the predicted faults obtained from dynamic information analysis as basic facts (\circled{4} in Fig.\ref{fig:method_fl4ml}). 
% All static information and predicted faults are incorporated into the KG as basic facts. 
Next,
KG rules are applied on the basic facts to infer fault-related facts and establish connections between all KG's facts, including basic and fault-related facts. 
% \Amin{I don't understand:}
Fault-related facts refer to the faults identified by \fltool{}. 
% \Amin{please reword this sentence to make it more clear, even divide it into 2 sentences}
Besides, a relationship between a fault-related fact and a basic fact referring to a part of the DL-based system indicates the location of the fault's root cause. 
For example, if the KG generator detects that no activation function has been assigned to the model layers, it creates a fault-related fact indicating a `missing activation function' issue in the model structure and links it to the fact representing the DL model.

\begin{table}[!t]
\caption{Statistical operators used to analyze extracted dynamic information from model training logs}
    \centering
    \begin{tabular}{p{2cm} l}
        \hline
            \textbf{Operator} & \textbf{Description} \\
            \hline
            min & The minimum value in a feature trace\\
            max & The maximum value in a feature trace\\
            median & The median value of a feature trace\\
            mean & The mean value of a feature trace\\
            var & The variance of a feature trace \\
            std & The standard deviation of a feature trace \\
            skew & The skewness of a feature trace\\
            sem & The standard error of the mean of a feature trace\\
        \hline
    \end{tabular}
        \label{tab:statistical_operators}
\vspace{-1em}
\end{table}

To derive conclusions from the KG and identify potential faults and their root causes in the DL-based system under test, a reasoning engine is required (\circled{7} in Fig.\ref{fig:method_fl4ml}). 
% \Amin{vel kon baba! I would remove this sentence:}The reasoning engine is an Artificial Intelligence (AI) system that applies various inference algorithms to process and infer insights from the existing knowledge~\cite{duan2020machine}.
The primary goal of the reasoning engine is to simulate rational reasoning, allowing machines to perform complex tasks such as problem-solving, decision-making, and more~\cite{bottou2014machine,duan2020machine}. For instance, a reasoning engine can be used to determine whether a patient has a specific disease by analyzing symptoms and historical health information.

% \Amin{you may put one or some algorithms to clearly mention the steps, and avoid mixing with implementation details}
\begin{table}[!t]
\caption{Fault that can be detected by \fltool{}}
    \centering
    \resizebox{\columnwidth}{!}{
    \small
    \begin{tabular}{p{4cm} p{10cm}}
        \hline
            \textbf{\textit{Fault}} & \textbf{\textit{Description}} \\
            \hline
            \rowcolor{lightgray}
            suboptimal train/test ratio  & split of the dataset into train and test is not optimal \cite{joseph2022optimal,tan2021critical}\\
            % [0.2cm]
            missing preprocessing & data preprocessing which should be done is missing 
            \\
            \rowcolor{lightgray}
            \textit{Python} version mismatch & \textit{Python} versions used in the training and deployed environments are not matched~\cite{morovati2023bugs}
            \\
            % [0.2cm]
            
            system architecture mismatch & CPU architecture of the systems used in the training and deployed environments are not matched~\cite{chen2020comprehensive}\\
            % [0.2cm]
            \rowcolor{lightgray}
            OS mismatch &  Operating System (OS) used in the training and deployed environments are not matched~\cite{chen2020comprehensive}\\
            % [0.2cm]
            
            libraries mismatch  &  version of installed libraries and frameworks mismatch in the training and deployed environments~\cite{morovati2023bugs}\\
            % [0.2cm]
            \rowcolor{lightgray}
            redundant activations & multiple and redundant connected activations can restrict the final activation from utilizing its full output range~\cite{nwankpa2018activation}\\
            % [0.2cm]
            % layers compatibility & A layer operating on an N-dimensional tensor should receive an input tensor with an exact N-dimensional shape~\cite{}\\[0.7cm]
            
            biases initialization & it is preferred to initialize biases to zero~\cite{glorot2010understanding}\\
            % [0.2cm]
            \rowcolor{lightgray}
            units initialization & the weight initialization should not be constant, as it is vital to break the symmetry between neurons~\cite{glorot2010understanding}\\
            % [0.2cm]
            
            non-linear activation & the activation function for learning layers, such as convolutional and fully-connected layers should be non-linear~\cite{nwankpa2018activation}\\
            % [0.2cm]
            \rowcolor{lightgray}
            loss linkage & the loss function should be properly defined and linked to the final layer's activation~\cite{nikanjam2021automatic}\\
            % [0.2cm]
            
            probability conversion & the final layer should include an activation function to convert the logits into probabilities for classification tasks~\cite{nikanjam2021automatic}\\
            % [0.2cm]
            \rowcolor{lightgray}
            suboptimal optimizer & the optimizer should be properly defined and integrated into the computational graph~\cite{nikanjam2021automatic}\\
            % [0.2cm]
            
            insufficient iteration & number of epochs is inadequate to reach the best model accuracy~\cite{cao2022deepfd}\\
            % [0.2cm]
            \rowcolor{lightgray}
            suboptimal learning rate & learning rate is insufficient for achieving good accuracy~\cite{cao2022deepfd}\\
            % [0.2cm]
            
            loss \& activation functions mismatch & loss and activation functions should be matched based on the model structure and data\\

            % \rowcolor{lightgray}
            % neurons suspension & the dropout layer should be positioned after the maximum pooling layer to enhance its effectiveness~\cite{srivastava2014dropout}\\
            % [0.2cm]
            
            % useless bias & learning layers should exclude bias when they are followed by batch normalization\\
            % [0.2cm]
            
            % \rowcolor{lightgray}
            % representative estimation & Batchnorm layer should be followed by dropout~\cite{nikanjam2021automatic}\\
            
            % layers compatibility & a processing layer that operates on a tensor must receive a valid input tensor with the exact expected dimensional shape \\
            \rowcolor{lightgray}
            valid intermediate layer& 
            the intermediate output of the layers should not be ``None'' or any similar values\\
            
            wrong activation function & 
            activation function should be defined based on the structure of the data\\
            
            \rowcolor{lightgray}
            missing activation function & 
            all models need activation functions in their layers to be able to learn patterns\\

        \hline
    \end{tabular}
    }
    \label{tab:fl4ml_rules_reasoning}
\end{table}

\begin{figure}[!b]
    \centering
    \includegraphics[width=0.8\linewidth]{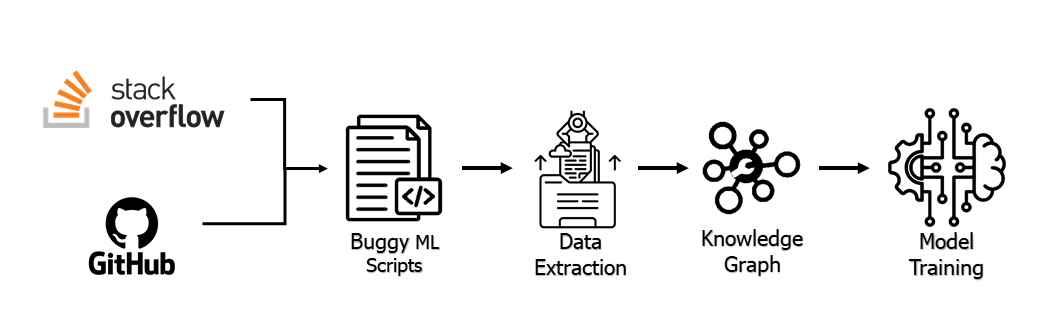}
    \caption{Training a model for KG link prediction}
    \label{fig:kg_link_predict}
\end{figure}

\subsection{Predicting Missed Relationship in KG}
Link prediction is a fundamental task for KGs used to complete relationships among nodes~\cite{nickel2015review}. \fltool{} leverages inductive link prediction to identify and add possible missing relationships within the generated KG (presented as \circled{5} in Fig.\ref{fig:method_fl4ml}). 
For this purpose, \fltool{} employs NodePiece~\cite{galkin2021nodepiece}, one of the latest and most efficient algorithms for link prediction in KGs.
NodePiece is an anchor-based method that learns a connected multi-relational graph by generating a combinatorial number of sequences based on the various types of relationships present in the KG. To train the model using NodePiece,
% (presented as \circled{5} in Fig.\ref{fig:method_fl4ml}), 
we utilize the training dataset, detailed in Subsection \ref{subsec:dataset}. 
Fig. \ref{fig:kg_link_predict} represents an overall view of the approach we use to train NodePiece.

\subsection{Ranking the Identified Root Causes}
Since \fltool{} may detect multiple faults during the localization process, we rank the identified faults to help DL developers prioritize the most probable root causes (presented as \circled{8} in Fig.\ref{fig:method_fl4ml}). To achieve this, we base the ranking on the relative frequency 
% \Amin{relative frequency?} 
of various faults reported in previous studies~\cite{chen2020comprehensive,humbatova2020taxonomy}. 
% \Amin{do you want to justify that we have different frequencies? if it's the case just say that! like: it's an example, but example of what?!}
That is, according to Humbatova et al.~\cite{humbatova2020taxonomy} that found faults related to the loss function (accounting for 11.7\% of training faults) are more common than those caused by the model optimizer (which account for 3\% of training faults), \fltool{} orders loss function faults as more likely than optimizer-related faults. 

\begin{algorithm}[t]
\caption{Localization of faults using generated KG}\label{alg:fl4ml_faults_localize}
\renewcommand{\algorithmicrequire}{\textbf{Input:}}
\renewcommand{\algorithmicensure}{\textbf{Output:}}
\begin{algorithmic}[]
\small
\rmfamily

\Require Extracted data (static and dynamic)
\Ensure Ranked list of fault's root causes

\State $data \gets predictedFaults + statitcInfo$
\State $KG \gets$ KgGenerator$(data)$ 
\newline \color{black!75} \ttfamily \hspace{1cm} // Generating a KG based on the extracted and analyzed info

\color{black} \rmfamily
\State $updatedKG \gets$KgLinkPrediction$(KG)$
\newline \color{black!75} \ttfamily \hspace{1cm} // Predicting missed KG's relationship using NodePiece

\color{black} \rmfamily
\State $RCs \gets$ reasoningEngine$(updatedKG)$
\newline \color{black!75} \ttfamily \hspace{1cm} // Localizing faults by reasoning on the created KG

\color{black} \rmfamily
\State $RankedRCs \gets$ rank$(RCs)$
\newline \color{black!75} \ttfamily \hspace{1cm} //Ranking identified faults based on their frequency

\end{algorithmic}
\end{algorithm}

\section{Results and Analysis}
\label{sec:result}

This section presents and discusses the findings of this study, alongside a detailed comparison of \fltool{} performance with other state-of-the-art fault localization approaches for DL-based systems. To this end, we aim to answer the following Research Questions (RQs): \\
\textit{\textbf{RQ1.} \textbf{[Evaluation]} How does \fltool{} compare against other fault localization approaches for DL-based systems?}\\
\textit{\textbf{RQ2.} \textbf{[Sensitivity Analysis]} To what extent does each component contribute to the overall performance of \fltool{}?}\\
\textit{\textbf{RQ3.} \textbf{[Error Analysis]} What are the characteristics of faults misclassified by \fltool{} in DL-based systems?}\\
All materials utilized to answer these RQs, such as the collected datasets and the source code of \fltool{}, are accessible in our replication package~\cite{replication_package_fl4ml}.

\begin{table}[!b]
\caption{Detailed information about the selected DL frameworks}
    \centering
    \small
    \begin{tabular}{p{3.5cm} r r r}
        \hline
            \textbf{DL Framework} & \textbf{\#stars} & \textbf{\#forks} & \textbf{\#subscribers} \\
            \hline
            \rowcolor{lightgray!50}
            TensorFlow & $174 k$ & $88.3 k$ & $3.4 k$\\
            % \hline
            \rowcolor{lightgray!50}
            Keras & $58.3 k$ & $19.3 k$ & $2 k$\\
            % \hline
            PyTorch & $66.7 k$ & $18.3 k$ & $1.7 k$ \\
            Caffe & $ 33.3 k$ & $19 k$ & $269$ \\
            Jax & $ 23.1 k$ & $2.2 k$ & $504$ \\
            MXNet & $ 20.4 k$ & $6.9 k$ & $875$ \\
            CNTK & $ 17.4 k$ & $4.4 k$ & $201$ \\
            Sonnet & $ 9.6 k$ & $1.4 k$ & $51$ \\
        \hline
    \end{tabular}
        \label{tab:ml_frameworks}
\end{table}

\subsection{Experimental Design}

Given the widespread popularity of \textit{Keras} and \textit{TensorFlow}, as shown by the metrics in Table~\ref{tab:ml_frameworks}, our focus in this study is on DL software systems developed using these two leading frameworks. 
To extract the static and dynamic information required by \fltool{}, we implement three main components to gather data related to the dataset, model training and its environment, and system deployment. Listing \ref{list_fltool_sample_usage} provides a sample of how \fltool{} can be used within the code of a DL-based application.
As demonstrated, integrating \fltool{}'s APIs into DL-based applications is straightforward and does not require specialized expertise in DL or its development. 
To collect data-related information, \fltool{} extracts various dataset details such as the number of features, the number of rows in the training set, the number of rows in the test set, etc. (illustrated as \circled{1} in Listing \ref{list_fltool_sample_usage}).
For static information (such as model structure, layers, activation functions, loss function, optimizer, etc.) and dynamic information related to the DL model and its training environment, we implement a \textit{Keras} callback function (illustrated as \circled{2} in Listing \ref{list_fltool_sample_usage}). 
The dynamic information collected includes neuron weights, accuracy, loss, validation accuracy, and validation loss, which are recorded at the end of each training epoch.

\begin{lstlisting}[language=Python, float, floatplacement=H, label=list_fltool_sample_usage,escapechar=\%,
caption=An example usage of \fltool{} within a DL code (highlighted lines are related to \fltool{}),
]
(x_train, y_train), (x_test, y_test) = keras.datasets.mnist.load_data()
%\Hilight%data_analysis(train=x_train, target=y_train, test=x_test) %\textbf{\circled{1}}%

model = keras.Sequential([
            keras.Input(shape=input_shape),
            ...
            layers.Dense(num_classes, activation="softmax")
        ])
model.compile(loss="categorical_crossentropy", 
            optimizer="adam",metrics=["accuracy"])
model.fit(x_train, y_train, batch_size=128, epochs=10, validation_split=0.1,
    callbacks=[
%\Hilight%      fl4ml(batch = 128, epochs = 10 ,data=[x_train, y_train, x_test, y_test]) %\textbf{\circled{2}}%
    ])
\end{lstlisting}

To generate KG from the extracted information, we use RDFLib~\cite{rdflib_repo}, a Python library for the Resource Description Framework (RDF). RDF provides a foundation for decision-making, moving beyond a system of locally trusted facts. Moreover, we utilize Notation3 (N3)~\cite{notation3}, a logic language known as a superset of RDF to implement KG. 
N3 generally enhances RDF by extending its representational capabilities and enabling decision-making through operations on data, information access, and reasoning. Besides, to infer the possible faults and their root causes from the generated KG, we use the `Euler Yet another proof Engine' (EYE)~\cite{verborgh2015drawing} as the reasoning engine of \fltool{}. EYE reasoning engine supports RDF and implements N3. 
EYE is notably more expressive and significantly outperforms other reasoning engines~\cite{de2015event}. For example, EYE can solve the Deep Taxonomy Benchmark problem with 100,000 triples in just 4.8 seconds, compared to the CWM reasoner, which takes 9 days and faces out-of-memory issues when using Jena~\cite{coom2002, jena}.
Moreover, to implement NodePiece which is used by \fltool{} to train a model for predicting possible missed relationships in the created KG, we have used PyKeen~\cite{ali2021pykeen} \textit{Python} library. 
% It is worth mentioning that PyKeen~\cite{ali2021pykeen}, a Python package to train and evaluate KG embedding models, 
% which has implemented NodePiece~\cite{galkin2022nodepiece} is used in our proposed approach. 

\begin{table}[!b]
 \caption{Sample results of comparing \fltool{}, and other popular fault localization techniques for DL-based systems}
    \begin{adjustbox}{max width=\textwidth}
    \centering	
    % \resizebox{\columnwidth}{!}{
    \begin{tabular}{l p{7.5cm} p{3cm} p{1.7cm} p{4.7cm} p{4.3cm}}
        \hline
            \multirow{ 2}{*}{Ref \#} & \multicolumn{5}{c}{Fault localization techniques}\\
            \cline{2-6}
             & \multicolumn{1}{c}{\textit{UMLAUT}} & 
             \multicolumn{1}{c}{\textit{DeepFD}} & 
             \multicolumn{1}{c}{\textit{AutoTrainer}} & 
             \multicolumn{1}{c}{\textit{DeepLocalize}}
             & 
             \multicolumn{1}{c}{\textbf{\fltool{}}} \\ 
             \hline
             \rowcolor{lightgray}
             \multirow{2}{*}{34311586} & 
             {
                 (1) Critical: Missing Softmax layer before loss \newline
                 (2) Warning: Last model layer has nonlinear activation
             } 
             &
             1:[lr] (Lines:27) 
             & \multirow{2}{*}{--} & 
             {
                Batch 0 layer 2: Error in Weights, \newline
                terminating training
             }
             & 
             1. suboptimal learning rate
             \newline
             2. wrong activation function
             \\
             \multirow{6}{*}{37624102} & 
             {
                (1) Critical: Missing Softmax layer before loss \newline
                (2) Critical: Missing activation functions \newline
                (3) Warning: Last model layer has nonlinear activation \newline
                (4) Error: Image data may have incorrect shape \newline
                (5) Warning: Learning Rate is high \newline
                (6) Warning: Check validation accuracy
             }& 
             {
                1:[lr] (Lines:66) \newline
                2:[Act] (Lines:54, 56, 61, 64)
             }
             &
             unstable & 
             {
                Batch 0 layer 9: Error in Output Gradient, \newline
                terminating training
             }
             &
             {
                1. suboptimal learning rate\newline
                2. wrong activation function
             }
             \\
             \rowcolor{lightgray}
             \multirow{3}{*}{41600519} & 
             {
                (1) Error: Input data exceeds typical limits \newline
                (2) Critical: Missing Softmax layer before loss \newline
                (3) Warning: Last model layer has nonlinear activation
             }
             & 
             1:[loss] (Lines:32) &
             unstable & 
             {
                Batch 0 layer 6: Error in forward \newline
                terminating training
             }
             & 
             1. loss linkage
             \\
             \multirow{2}{*}{47352366} & 
             {
                 (1) Critical: Missing Softmax layer before loss \newline
                 (2) Warning: Last model layer has nonlinear activation
             }
             & 
             1: [opt] (Lines:40)& 
             explode & 
             {
                Layer-12 Error in delta weights \newline 
                Stop at epoch 1, batch 24 
             }
             & 
             {
                1. loss \& activation function mismatch \newline
                2. suboptimal optimizer
             }
             \\ 
             \rowcolor{lightgray}
             \multirow{2}{*}{48385830} &
             {
                 (1) Critical: Missing Softmax layer before loss \newline 
                 (2) Warning: Possible overfitting
             }
             &
             {
                1:[act] (Lines:-) \newline
                2:[loss] (Lines:57)
             }
             &
             explode &
             {
                 Layer-1 Error in forward \newline 
                 Stop at epoch 1, batch 2
             }
             & 
             {
                1. missing activation function \newline
                2. loss linkage
             }
             \\
             
             \multirow{3}{*}{50079585} & 
             {
                 (1) Critical: Missing Softmax layer before loss \newline
                 (2) Critical: Missing activation functions \newline
                 (3) Warning: Last model layer has nonlinear activation
             }
             & 
             {
                1:[lr] (Lines:44) \newline
                2:[epoch] (Lines:15)
             }
             &
             unstable & 
             \multirow{3}{*}{--} 
             &  
             {
                1. suboptimal learning rate \newline
                2. insufficient iteration
             }
             \\
             \rowcolor{lightgray}
             \multirow{6}{*}{55328966}&
             {
                 (1) Error: Input data exceeds typical limits \newline
                 (2) Warning: Possible overfitting \newline
                 (3) Warning: Check validation accuracy \newline
                 (4) Critical: Missing Softmax layer before loss \newline
                 (5) Critical: Missing activation functions \newline
                 (6) Warning: Last model layer has nonlinear activation
             }
             & 
             1:[opt] (Lines:49)&
             explode & 
             \multirow{6}{*}{--} &  
             {
                1. loss \& activation functions mismatch \newline
                2. suboptimal optimizer
             }
             \\
             
             \multirow{3}{*}{59282996} & 
             {
                (1) Error: Input data exceeds typical limits \newline
                (2) Warning: Check validation accuracy \newline
                (3) Critical: Missing Softmax layer before loss
             }
             & 
             1:[epoch] (Lines:309) & 
             unstable & 
             \multirow{3}{*}{--} & 
             {
             1. suboptimal optimizer \newline
             2. wrong activation function
             }
             \\
             \hline
    \end{tabular}
        % }   
        \end{adjustbox}
       \label{tab:fl4ml_sample_result}
\vspace{-1em}
\end{table}

% \Amin{before this, define RQs, like: To evaluate \fltool{} on ... we define these RQs: ...}

% \subsection{RQ1. How effective is \fltool{} in localizing faults in DL-based systems?}

\subsection{RQ1 [Evaluation]}

To assess the effectiveness of our proposed approach, we use our validation dataset that includes 100 buggy DL samples
to compare our approach with previous approaches, including \textit{DeepFD} \cite{cao2022deepfd}, \textit{DeepLocalize} \cite{wardat2021deeplocalize}, \textit{AutoTrainer} \cite{zhang2021autotrainer}, and \textit{UMLAUT} \cite{schoop2021umlaut}. 
To run these approaches, we use their related replication packages that are publicly available. 
% \Amin{were there any particular settings to run these tools? if so, mention them}
Since our evaluation is based on 100 buggy DL samples, 
a sample set of gathered results using various approaches is presented in 
Table \ref{tab:fl4ml_sample_result}. 
% shows the result of various approaches for a sample set of buggy scripts
% \Amin{first say that since we have 100 samples, we provide a sample of results in Table 4 and table 5 reports all results, then we discuss results in details in the rest of this section}. 
% This resulted in a total of 100 samples 
Besides, Table \ref{tab:fl4ml_results} presents the number of faults identified by each approach.
% \Amin{tool?}.

Regarding data-related issues in training DL models, only \textit{UMLAUT} and \fltool{} successfully identify the associated faults. While \textit{UMLAUT} successfully localizes 42\% of data-related faults in DL software systems, \fltool{} outperforms it by accurately identifying and localizing 84\% of these faults. For faults caused by mismatches between installed libraries in the training and deployment environments, \fltool{} detects 100\% of such issues, owing to specific rules implemented for localizing deployment faults. In contrast, none of the other approaches from previous studies are capable of identifying these issues.
Faults related to loss and activation functions are among the most frequent in DL software systems~\cite{humbatova2020taxonomy}, and all studied approaches can detect them to varying degrees of accuracy. 
For loss function-related faults, \fltool{} and \textit{DeepLocalize} achieve the best performance, detecting 69\% and 63\% of these issues, respectively. In terms of activation function faults, \textit{UMLAUT} leads with 100\% accuracy, followed by \fltool{} with 92\%, and \textit{AutoTrainer} with 38\%.
Another significant source of faults in DL-based systems is the optimization function. Regarding optimization-related faults, \textit{DeepFD} with 50\%, \fltool{} with 30\%, and \textit{DeepLocalize} with 10\% have the best accuracy. Lastly, faults caused by an insufficient number of training iterations are only detected by \textit{DeepFD} and \fltool{}, with 60\% and 53\% detection rates, respectively.

Although \fltool{} does not achieve the best performance for all fault types, it demonstrates the most consistent overall performance among all approaches. Specifically, \fltool{} strikes a balance in identifying and localizing a wide range of DL faults. For instance, while \textit{UMLAUT} outperforms \fltool{} in identifying faults related to activation functions, its performance for other fault types is relatively low compared to other approaches. Similarly, \textit{DeepFD}, which excels in localizing faults related to insufficient iterations and optimization functions, performs poorly for other types of faults. In summary, while \fltool{} leads in three out of the six examined DL fault types, it ranks as the second-best approach for the remaining fault types. In contrast, other methods tend to perform well for only one fault type and show significantly weaker performance across the rest.

% \Amin{here, you can have a paragraph to summarize results, discussing pros/cons of all tools}
\begin{tcolorbox}
\textbf{Finding 1.} Based on the accuracy of identifying and localizing DL-related faults, \fltool{} outperforms other techniques for faults in data (84\%), loss function (69\%), and mismatch between installed libraries on the training and deployment environment (100\%).  
\end{tcolorbox}

\begin{table}[!t]
    \centering
    \caption{Results of comparing the proposed method with previous studies }
    \resizebox{\columnwidth}{!}{
    \begin{tabular}{p{4.5cm} r r r r r r}
        \hline
         \textbf{Issue Type} & \textbf{\# samples} & \textbf{\textit{DeepFD}} & \textbf{\textit{DeepLocalize}} & \textbf{\textit{AutoTrainer}} & \textbf{\textit{UMLAUT}} & \textbf{\textit{\fltool{}}} \\
         \hline
         Data                           & 19 & 0 & 0 & 0 & 8 & 16 \\
         Mismatch framework/libraries   & 20 & 0 & 0 & 0 & 0 & 20 \\
         Loss function                  & 16 & 5 & 10& 5 & 0 & 11 \\
         Insufficient iteration         & 15 & 9 & 0 & 0 & 0 & 8  \\
         Optimization function          & 10  & 5 & 1 & 0 & 0 & 3  \\
         Activation function            & 26 & 2 & 2 & 10& 26 & 24 \\
         \hline
    \end{tabular}
    }
    \label{tab:fl4ml_results}
\vspace{-1em}
\end{table}

\subsection{RQ2 [Sensitivity Analysis]}

This section highlights the contribution of each component of \fltool{} to the overall approach through an ablation study. Using the same 20 samples we used to compare approaches,
% methodologies\Amin{do we have consistent terminology? you always call them methodologies?}, 
we assess the sensitivity of \fltool{} to its components~\cite{meyes2019ablation}. The core idea of an ablation study is to systematically remove specific components during each execution and compare the results against a baseline. In this context, we establish the baseline by running \fltool{} with all components. 
Since \fltool{} relies on static information, dynamic information, and KG link prediction as its key components, we designed the ablation study to evaluate the impact of removing each component in separate execution scenarios.
Table ~\ref{tab:fl4ml_ablation_results} presents the results of executing \fltool{} under different conditions, showing the effect of removing various components on overall performance.

\begin{table}[!t]
    \centering
    \small
    \caption{Results of \fltool{} sensitivity analysis by removing Static Information (SI), Dynamic Information (DI), and KG Link Prediction (LP) components}
    \resizebox{0.85\columnwidth}{!}{
    \begin{tabular}{p{4.5cm} p{1.5cm} p{1.5cm} p{1cm} p{1cm} l}
        \hline
         \multirow{2}{*}{\textbf{Issue Type}} & 
         \multirow{2}{*}{\textbf{\# samples}}&  
         \multirow{2}{*}{\textbf{{\fltool{}}}}&
        \multicolumn{3}{c}{Removed Component} \\
        \cline{4-6}
        
         & & & \textbf{\textit{SI}} & \textbf{\textit{DI}} & \textbf{\textit{LP}} \\
         \hline
         Data                           & 19 & 16 & 5 & 16 & 15  \\
         Mismatch framework/libraries   & 20 & 20  & 0 & 20 & 20  \\
         Loss function                  & 16 & 11 & 5 & 8 & 9  \\
         Insufficient iteration         & 15 & 8  & 8 & 2 & 6  \\
         Optimization function          & 10 & 3  & 3 & 1 & 2  \\
         Activation function            & 26 & 24 & 6 & 22 & 22  \\
         \hline
    \end{tabular}
    }
    \label{tab:fl4ml_ablation_results}
\end{table}

As the results demonstrate, removing static information has the most significant impact on the performance of \fltool{}, leading to a 67\% performance reduction. This substantial decrease is due to the fact that static information is crucial for identifying a wider range of fault types compared to other components. For instance, data-related faults can be detected through the use of extracted static information.
Besides, the number of faults that their identification and localization highly depend on static information is higher than ones relying on dynamic information. 
Dynamic information and KG link prediction follow with 16\% and 10\% impacts, respectively. The relatively lower influence of the KG link prediction component may be attributed to the limited size of the training dataset, as dataset size has a direct effect on model performance~\cite{sun2017revisiting}.
To address this, we plan to enrich our training dataset as a future work, which is expected to improve the efficiency of the KG link prediction component in \fltool{}.

For statistical analysis of the results, %To assess the statistically significant differences in the performance of \fltool{} across baseline and other scenarios\Amin{I would say:  ...}, 
we employed Fisher's exact test~\cite{klein2006survival}. This choice is justified by our aim to evaluate the statistical significance between pairs of values (e.g., baseline and SI) given our sample size of approximately 100. %, Fisher’s exact test is the suitable choice.
% considered appropriate\Amin{a feasible statistical test?}
~\cite{arcuri2014hitchhiker}. We leverage the SciPy Python library~\cite{2020SciPy-NMeth} to conduct the test. 
The results show a significant difference in \fltool{}'s performance when the static information component is removed (\textit{p-value} = 0.04). However, for the scenarios involving dynamic information and KG link prediction removal, no significant differences are found, with \textit{p-values} of 0.06 and 0.10, respectively. It is worth noting that \textit{p-values} between 0.05 and 0.10 are interpreted as marginal significance, meaning that the evidence against the null hypothesis is weak but not entirely negligible~\cite{hackshaw2011interpreting}.

\begin{tcolorbox}
\textbf{Finding 2.} 
Static information makes the most substantial contribution to \fltool{}'s performance, with its removal causing a 67\% performance drop and showing a statistically significant difference. In contrast, KG link prediction has the least impact on \fltool{}'s performance, possibly due to the small size of the training dataset

%Static information contributes the most substantially to the performance of \fltool{}, as its removal results in a 67\% performance drop with a statistically significant difference. In contrast, KG link prediction has the least impact on \fltool{}'s performance, though this lower contribution may be due to the small size of the training dataset.
\end{tcolorbox}

\subsection{RQ3 [Error Analysis]}

Given that precision and recall are widely used metrics for comparing the effectiveness of different methods~\cite{goutte2005probabilistic}, we also assess the performance of the studied approaches using these metrics. Table \ref{tab:fl4ml_metrics} presents the \textit{False Positives} (FP), \textit{False Negatives} (FN), \textit{Precision} (PR), and \textit{Recall} (RC) for identifying various ML-related faults.

For data-related faults in DL-based systems, \fltool{} and  \textit{UMLAUT} are the only ones capable of detecting such faults. \fltool{} achieves the highest performance with precision and recall values of 100\% and 84\%, respectively in detecting data-related faults, while \textit{UMLAUT} stays behind with precision and recall values of 23\% and 44\%, respectively. 
It is also worth mentioning that we do not observe any FP for \fltool{}, where the results of \textit{UMLAUT} show 27 FP. 
In terms of identifying faults in the loss function, \fltool{}, \textit{DeepFD}, and \textit{DeepLocalize} outperform other methods. 
Notably, although \textit{DeepFD} and \textit{DeepLocalize} exhibit opposite trends in precision and recall, \fltool{} performs better on both metrics (with 85\% for precision and 69\% for recall). In other words, \fltool{} with only 2 FP in localizing faults regarding loss function shows better performance. 
When it comes to issues caused by insufficient training iterations, \textit{DeepFD} and \fltool{} demonstrate nearly identical performance, with \fltool{} showing slightly better results, in terms of both precision and recall. Besides, \fltool{} performs better concerning 1 FP, in comparison with 2 FP for \textit{DeepFD}. 
For detecting and localizing issues in the optimization function, \textit{DeepFD} performs the best, with precision and recall of 36\% and 50\%, respectively, while \fltool{} ranks second.
However, it should be taken into account that \fltool{} results in fewer FP for localizing faults regarding optimization functions. 
For faults related to the activation function, although \textit{UMLAUT} achieves the highest recall, its precision is just 26\%, significantly lower than its recall. This large gap between precision and recall is due to \textit{UMLAUT}'s tendency to report activation issues for all tested faults. \textit{AutoTrainer} also identifies activation function issues, achieving a precision of 91\%, but it's recall is 38\%, again showing a significant disparity between the two metrics. In contrast, \fltool{} stands out as the best-performing method in this category, with a well-balanced precision of 89\% and recall of 92\% for localizing activation function faults. Besides, \fltool{} with 3 FP in localizing faults related to the activation function stays at the second rank,
% \Foutse{you mean second position? rank?}, 
with respect to the number of FP.

For data-related faults in DL-based systems, \fltool{} and \textit{UMLAUT} are the only methods capable of detecting such faults. \fltool{} delivers the best performance, achieving precision and recall rates of 100\% and 84\%, respectively, while \textit{UMLAUT} lags behind with precision and recall values of 23\% and 44\%, respectively. Notably, \fltool{} produces no FP, whereas \textit{UMLAUT} records 27 FPs.
In terms of identifying faults in the loss function, \fltool{}, \textit{DeepFD}, and \textit{DeepLocalize} outperform other approaches. Although \textit{DeepFD} and \textit{DeepLocalize} display opposite trends in precision and recall, \fltool{} surpasses both, with 85\% precision and 69\% recall, and only 2 FPs for loss function-related faults.
For issues arising from insufficient training iterations, both \textit{DeepFD} and \fltool{} demonstrate similar performance, though \fltool{} slightly edges out in both precision and recall. Additionally, \fltool{} gives only 1 FP, compared to 2 FPs for \textit{DeepFD}.
When it comes to detecting and localizing issues in the optimization function, \textit{DeepFD} performs the best, with precision and recall values of 36\% and 50\%, respectively, while \fltool{} ranks second. However, \fltool{} produces fewer FPs in localizing faults related to the optimization function.
For faults related to the activation function, while \textit{UMLAUT} achieves the highest recall, its precision is just 26\%, significantly lower than its recall. This disparity stems from \textit{UMLAUT}'s tendency to flag activation issues for all tested faults. \textit{AutoTrainer} also detects activation function issues with 91\% precision, but its recall is only 38\%, indicating a notable gap between the two metrics. In contrast, \fltool{} emerges as the top-performing method in this category, with balanced precision and recall values of 89\% and 92\%, respectively. Additionally, \fltool{}, with 3 FPs in localizing activation function faults, ranks second in terms of the number of FPs.

% In terms of FP, 
% \Amin{discuss FP explicitly!}

\begin{table}[!t]
    \centering
    \caption{Comparison of False Positive (FP), False Negative (FN), Precision (PR), and Recall (RC) of various approaches in localizing ML-related faults}
     \resizebox{\columnwidth}{!}{
        \begin{tabular}{p{4.5cm} r r r r | r r r r | r r r r | r r r r | r r r r }
            \hline
            \multirow{2}{*}{\textbf{Issue Type}} & \multicolumn{4}{c|}{\textit{\textbf{DeepFD}}} & \multicolumn{4}{c|}{\textit{\textbf{DeepLocalize}}} & \multicolumn{4}{c|}{\textbf{AutoTrainer}} & \multicolumn{4}{c|}{\textbf{UMLAUT}} & \multicolumn{4}{c}{\textbf{\fltool{}}} \\
            
            & FP & FN & PR & RC & FP & FN & PR & RC & FP & FN & PR & RC & FP & FN & PR & RC & FP & FN & PR & RC \\
            \hline
            \rowcolor{lightgray}
            Data & 0 & 19 & 0 & 0 & 0 & 19 & 0 & 0 & 0 & 19 & 0 & 0 & 27 & 10 & 0.23 & 0.44 & 0& 3 & 1.00 & 0.84 \\
            Mismatch framework/libraries & 0 & 20 & 0 & 0 & 0 & 20 & 0 & 0 & 0 & 20 & 0 & 0 & 0 & 20 & 0 & 0 & 0 & 0 & 1.00 & 1.00 \\
            \rowcolor{lightgray}
            Loss function & 3 & 11 & 0.63 & 0.31 & 20 & 6 & 0.33 & 0.63 & 18 & 11 & 0.22 & 0.31 & 0 & 16 & 0 & 0 & 2 & 5 & 0.85 & 0.69 \\
           Insufficient iteration & 2 & 6 & 0.82 & 0.60 & 0 & 15 & 0 & 0 & 0 & 15 & 0 & 0 & 0 & 15 & 0 & 0 & 1 & 5 & 0.89 & 0.62 \\
           \rowcolor{lightgray}
           Optimization function & 9 & 5 & 0.36 & 0.50 & 15 & 9 & 0.06 & 0.10 & 0 & 10 & 0 & 0 & 8 & 10 & 0 & 0 & 7 & 7 & 0.30 & 0.30 \\
           Activation function & 3 & 24 & 0.40 & 0.08 & 18 & 24 & 0.10 & 0.08 & 1 & 16 & 0.91 & 0.38 & 74 & 0 & 0.26 & 1.00 & 3 & 2 & 0.89 & 0.92 \\ 
            
            \hline
        \end{tabular}
    }
    \label{tab:fl4ml_metrics}
\vspace{-1em}
\end{table}

\begin{tcolorbox}
\textbf{Finding 3.} According to the \textit{Precision} and \textit{Recall} of localizing DL faults, \fltool{} has better performance for issues related to the data, mismatch between installed libraries on the training and deployment environments, loss function and insufficient training iterations. Although regarding faults within activation functions \textit{AutoTrainer} and \textit{UMLAUT} have the best precision and recall respectively, \fltool{} has better performance with respect to the balance between precision and recall.
Besides, \fltool{} produces fewer FP in localizing all types of faults, except faults regarding the activation function that its number of recorded FP stays at the second place. 
% \Amin{discuss FP explicitly!} 
\end{tcolorbox}

As the results indicate, \textit{DeepFD} outperforms \fltool{} in identifying faults related to `insufficient iterations' and `optimization function'. This difference can largely be attributed to the fact that \textit{DeepFD} executes the buggy scripts 10 times, whereas \fltool{} relies on a single execution. 
It is well-acknowledged that faults related to `insufficient iterations' and `optimization function' are best identified by analyzing and monitoring the dynamic information during model training~\cite{goodfellow2016deep,pascanu2013difficultytrainingRNN}. 
Given the inherent randomness in DL, it is reasonable that results based on multiple executions tend to be more accurate than those from a single execution. However, it is important to consider that running multiple training executions can be highly costly and time-consuming~\cite{panichella2021we}.
% \Amin{then the question would be why we don't have multiple runs, as you imply that we need to have!}
On the other hand, \textit{UMLAUT} shows the best performance in localizing faults related to the `activation function', with an 8\% higher accuracy than \fltool{}. However, this result can be explained by the fact that \textit{UMLAUT} always reports `activation function' faults for all tested scripts
% \Amin{don't understand, so it reports activation errors always?}. 
As such, this higher performance should not be interpreted as better fault localization. In fact, \textit{UMLAUT}'s very low precision (0.26\%) in identifying `activation function' faults confirms that its approach lacks accuracy for this type of fault.

\begin{tcolorbox}
\textbf{Finding 4.} Although \textit{DeepDF} outperforms \fltool{} in localizing faults related to `insufficient training iteration' and `optimization function', it is more computationally expensive in terms of execution time and resource consumption. 
Furthermore, while \textit{UMLAUT} surpasses \fltool{} in identifying faults in the 'activation function,' it consistently reports this fault for all tested DL codes, regardless of the presence of any fault in the `activation function'.
\end{tcolorbox}

\section{Threats to validity}
\label{sec:threats}

\subsection{Construct Validity}
A potential limitation of this study stems from the selection of buggy scripts used to train DL models, which were central to both our proposed approach and comparison with prior approaches. To select buggy DL code, we relied solely on real-world samples sourced from SO posts and GitHub repositories. Specifically, we used publicly available benchmarks of DL bugs~\cite{morovati2023bugs,cao2022deepfd,wardat2021deeplocalize} to extract relevant samples, and supplemented them with additional buggy scripts identified through a review of SO posts related to DL faults. To compare \fltool{} with existing fault localization approaches for DL-based systems, we used buggy DL scripts previously employed by other studies to demonstrate the effectiveness of their approaches. Moreover, we applied several mutation operators on them~\cite{chen2020comprehensive,humbatova2021deepcrime} to generate additional, unseen samples for a more comprehensive evaluation.
Furthermore, to minimize bias in the evaluation, we employed a set of buggy samples that had been used in previous approaches for comparison. 
% \Amin{you can also add that: what we did in this paper is similar to what others did in their papers, and cite them. also if we did anything to avoid bias and keep the comparison fair, ...}

\subsection{Internal Validity}
The first internal threat to the validity of this study is the limited number of buggy scripts used to train the ML models in various components of \fltool{}. This constraint arises because we only utilized real-world buggy scripts sourced from SO and GitHub.
Moreover, we focus solely on reproducible buggy samples, filtering out non-reproducible ones which further limits the size of the training dataset. 
Bug reproducibility is a well-known challenge in testing DL-based software systems, considering it has been reported that only about 3.34\% of DL bugs reported on SO are reproducible~\cite{morovati2023bugs,shah2024towards}.
% , which had to be implemented in \textit{Python} 3.8, \textit{TensorFlow} 2.2.0, and \textit{Keras} 2.3.1 (or newer versions). 
% Additionally, we focus exclusively on reproducible buggy samples.
% \Foutse{what do you mean here? i don't understand! are you trying to say you could not localize unreproducible bugs? be precise please!}. 
% It is worth mentioning that 
% bug reproducibility is considered a well-known challenge in testing DL software systems~\cite{shah2024towards}. For example, it is reported that only about 3.34\% of DL bugs reported in SO are reproducible~\cite{,morovati2023bugs}.
% —further contribute to this limitation\Amin{clearly say that: we only consider reproducible buggy samples ...}.
Another potential threat to internal validity is the selection of KNN, Random Forest (RF), and Decision Tree models for analyzing dynamic information. Although these are relatively simple ML models, they are efficient and widely used ~\cite{rigatti2017random,zhang2021challenges}. Moreover, given the limited amount of data available for training, these models are likely more effective than more complex alternatives.

\subsection{External Validity}
% The most significant threat to the external validity of \fltool{} is its restriction to ML applications developed using \textit{Keras} or \textit{TensorFlow}. These frameworks were chosen because they are the two most popular ML frameworks~\cite{top_ml_frameworks}. 
% Concerning that these frameworks are mostly used for DL applications' development, it is possible that 
% Another potential threat to the external validity of \fltool{} is the callback function specifically designed for \textit{TensorFlow 2.2.0} and \textit{Keras 2.3.1} or newer versions. These versions were selected as they were the first to be compatible with \textit{Python} 3.8, which was released in 2019~\cite{python38}. Consequently, older versions of \textit{Keras} and \textit{TensorFlow} are not compatible with \textit{Python} 3.8, further limiting the tool's applicability.
% Another potential threat to the external validity of this study is our focus on ML-based systems implemented using \textit{Python}. Given that \textit{Python} is the most common programming language for the development of ML applications~\cite{voskoglou2017best,Gupta:MLLangugae}, we believe that the results and findings of this study can be generalized to the majority of ML-based systems. 
The primary threat to the external validity of \fltool{} is its limitation to DL software systems developed using \textit{Keras} or \textit{TensorFlow}. These frameworks were selected due to their popularity as the two most widely used DL frameworks~\cite{top_ml_frameworks}. It is also worth mentioning that \fltool{} can be extended to other DL frameworks such as \textit{PyTorch}. 
% \Amin{say that \fltool{ can be extended to other libraries like PyTorch ...}}
Another potential threat to the external validity of this research is our focus on DL-based systems implemented using \textit{Python}. Given that \textit{Python} is the most commonly used language for developing DL software systems~\cite{voskoglou2017best,Gupta:MLLangugae}, we believe that the results and conclusions of this study can be generalized to the majority of DL-based systems.

\subsection{Reliability validity} 
We explained the methodology used in \fltool{} in detail and provided a replication package~\cite{replication_package_fl4ml} allowing others to reproduce our results and expand our proposed methodology. 

\section{Conclusion and Future Works}
\label{sec:conclusion}

In this paper, we introduced \fltool{}, a fault localization technique designed for DL-based systems. Unlike existing fault localization methodologies that primarily focus on the DL model and its training, \fltool{} takes a system-level approach, addressing the entire DL system pipeline. 
Additionally, \fltool{} leverages both static and dynamic information from DL-based systems to enhance the accuracy of fault identification and localization.
Our comparison of \fltool{} with four previously published fault localization techniques for DL-based systems demonstrates that \fltool{} outperforms these methods in three out of six fault categories, based on accuracy. 
Moreover, \fltool{} shows superior performance in four out of six fault types when evaluated by precision and recall.
For future work, we aim to expand our training dataset by incorporating more real-world buggy DL samples, thereby improving the performance of the ML models used within \fltool{}. 
Additionally, we plan to enrich the fact extraction step by gathering more comprehensive information and developing additional rules to support the identification and localization of a broader range of DL faults.

% \begin{acks}
%     This work was supported by: Fonds de Recherche du Québec (FRQ), the Canadian Institute for Advanced Research (CIFAR) as well as the DEEL project CRDPJ 537462-18 funded by the National Science and Engineering Research Council of Canada (NSERC) and the Consortium for Research and Innovation in Aerospace in Québec (CRIAQ), together with its industrial partners Thales Canada inc, Bell Textron Canada Limited, CAE inc and Bombardier inc.
% \end{acks}
% \bibliographystyle{IEEEtran}
% \bibliography{bibliography.bib}

\bibliographystyle{ACM-Reference-Format}
\bibliography{biblio.bib}

% \appendix









\end{document}